\documentclass[12pt, a4paper]{article}
\pdfoutput=1
 \usepackage[font=small,format=plain,labelfont=bf,up,textfont=normal,up,justification=justified,singlelinecheck=false]{caption}

\usepackage[a4paper, left=2cm,right=2cm]{geometry}
\usepackage[colorlinks=true,linkcolor=black,citecolor=teal,urlcolor=MidnightBlue,filecolor=black]{hyperref}
\usepackage{amsfonts}
\usepackage{amsmath}
\usepackage{setspace}
\usepackage[dvipsnames]{xcolor}
\definecolor{SchoolColor}{rgb}{0.6471, 0.1098, 0.1882} 

\usepackage[utf8,applemac]{inputenc}
\usepackage{tensor}
\usepackage{cite}
\usepackage{tikz}
\usepackage{graphicx}
\usepackage{bm} 

\usepackage{dcolumn}
\usepackage{bm}

\newcommand{\bea}{\begin{eqnarray}}
\newcommand{\eea}{\end{eqnarray}}
\newcommand{\be}{\begin{equation}}
\newcommand{\ee}{\end{equation}}
\def\nn{\nonumber}

\newcommand{\beqs}{\begin{eqnarray}}
\newcommand{\eeqs}{\end{eqnarray}}

\numberwithin{equation}{section}

\begin{document}
\begin{titlepage}

\begin{flushright}\vspace{-3cm}
{\small
\today }\end{flushright}
\vspace{0.5cm}
\begin{center}
	{{ \LARGE{\bf{Correlation function of modular  \vspace{6pt}\\ Hamiltonians }}}} \vspace{5mm}

	\centerline{\large{\bf{Jiang Long\footnote{e-mail:
				 longjiang@hust.edu.cn}}}}
	\vspace{2mm}
	\normalsize
	\bigskip\medskip
	\textit{Asia Pacific Center for Theoretical Physics,\\ Pohang 37673, Korea}\\\
\vspace{2mm}

	\textit{School of Physics, Huazhong University of Science and Technology, \\Wuhan, Hubei 430074, China
	}
	
	\vspace{25mm}
	
	\begin{abstract}
		\noindent
		{We investigate varies correlation functions of modular Hamiltonians defined with respect to spatial regions in quantum field theory. These correlation functions are divergent in general. We extract finite correlators by removing divergent terms for two dimensional massless free scalar theory. We reproduce the same correlators in general two dimensional conformal field theories.} 
	\end{abstract}
	

\end{center}

\end{titlepage}
\tableofcontents
\section{Introduction}
Modular Hamiltonian of a spatial subregion is  a natural and fundamental object in QFT\cite{Haag:1992}. In QFT defined with a cut-off, it is the logrithmic of  the reduced dentity matrix $\rho_A$, $\hat{H}_A=-\log\rho_A$. It plays a central role in quantum information theory. Entanglement entropy, relative entropy and  other important quantities are  constructed from modular Hamiltonian\cite{Araki:1976zv,Bombelli:1986rw,Srednicki:1993im,Callan:1994py}. It has also been used to obtain many interesting results, such as the Bousso bound \cite{Casini:2008cr,Bousso:2014sda,Bousso:2014uxa}, first law of entanglement entropy \cite{Blanco:2013joa}, the proof  of varies energy conditions \cite{Faulkner:2016mzt,Balakrishnan:2017bjg} and contrains of correlation functions\cite{Lashkari:2018nsl}.  

Modular Hamiltonian is in general a highly non-local operator, though for certain symmetric situations it could be analytic. When the subregion is Rindler wedge of Minkowski spacetime and the state is in vacuum, modular Hamitonian is the boost generator which is a smeared operator of stress tensor in Rindler wedge \cite{Bisognano:1976}. A nice realization of this result is Unruh effect \cite{Unruh:1976db}.  For a spherical region in a conformal field theory, an integral form could be found by conformal transformation of the Rindler wedge \cite{Casini:2011kv}. In two dimensional free field theory, a bilocal form of modular Hamiltonian can also be obtained for several disjoint intervals \cite{Casini:2009sr, Arias:2018tmw}. For a conformal field theory in a state which has a gravitational dual, a $1/G_N$ expansion of modular Hamiltonian has also been proposed \cite{Jafferis:2014lza, Jafferis:2015del} from bulk, 
\be
\hat{H}_{\text{bdy}}=\frac{\hat{A}_{\text{ext}}}{4 G_{N}}+\hat{H}_{\text{bulk}}+\cdots+o(G_{N}),\label{holoH}
\ee 
where the first term is an area operator of Ryu-Takayanagi\cite{Ryu:2006bv} surface, $\hat{H}_{\text{bulk}}$ and other higher order terms are bulk modular Hamiltonian of the corresponding bulk region. Unfortunately, modular Hamiltonian is not additive, in the sense that modular Hamitonian of two disjoint regions is in general not the summation of modular Hamiltonian in each seperate region 
\be
\hat{H}_{A\cup B}\not= \hat{H}_{A}+\hat{H}_B,\label{acupb}
\ee 
which makes it hard to study multi-region quantities. 

R\'enyi entropy is divergent for continuous quantum field theories \footnote{Recently, modular Hamiltonian is also discussed from algebraic quantum field theory \cite{Witten:2018lha}. The technical difficulty to define reduced density matrix in general continuous quantum field theory does not affect the result in this work. We will also not discuss the total modular Hamiltonian $\tilde{H}=H_A-H_{\bar{A}}$ as many other references.}, however, one can still obtain interesting information by taking care of the cut-off.  By definition, it is essentially a combination of correlation functions of modular Hamiltonians, the divergence is partly due to the singularity when two or more modular Hamiltonians collide in the same region. Therefore, the following correlation function 
\be
\langle \hat{H}^m_A\hat{H}^n_B\rangle,\quad m\ge1,n\ge1\label{mnhab}
\ee 
is also divergent in general. However, we will show that one can extract interesting finite result by removing the divergent part carefully in specific examples. The point is to define so called connected correlation function 
\be
T_{A\cup B}^{(m,n)}\equiv\langle \hat{H}_A^m\hat{H}_B^n\rangle_c,\quad m\ge 1,n\ge 1\label{deftmn}
\ee 
for two finite disjoint regions. We will specify the details of the construction and compute (\ref{deftmn}) for massless free scalar and general conformal field theories in  the following sections. The correlation function (\ref{deftmn}) is not directly related to R\'enyi entropy of two disjoint regions since (\ref{acupb}). However, it may provide another way to understand the correlation between two disjoint regions. 

The structure of this paper is as follows. We will introduce a generator of (\ref{deftmn}) in section 2 and study the generator explicity in two dimensinal massless free scalar theory in section 3. In section 4, we reproduce the same result of section 3 using technics of two dimensional conformal field theory. We will comment on the implications to operator product expansion of reduced density matrix and constraints on holograhic dual (\ref{holoH}) in section 5 and 6. Conclusions and discussions are collected in the last section. 

\section{A generator of correlation functions}
We consider a system with a density matrix $\rho$. The modular Hamiltonian corresponding to region $A$(or $B$) is $\hat{H}_A$ (or $\hat{H}_B$). $A$ and $B$ are assumed to be spacelike and disjoint, therefore from causality, the commutator of $\hat{H}_A$ and $\hat{H}_B$ is zero. A generator of correlation function (\ref{deftmn}) is 
\be
T_{A\cup B}(a,b)=\log\frac{\langle e^{-a\hat{H}_A-b\hat{H}_B}\rangle}{\langle e^{-a\hat{H}_A}\rangle\langle e^{-b\hat{H}_B}\rangle},\label{correlator}
\ee 
where we have inserted a normalization factor such that $T_{A\cup B}(a,b)$ is zero whenever $\langle e^{-a\hat{H}_A}e^{-b\hat{H}_B}\rangle=\langle e^{-a\hat{H}_A}\rangle\langle e^{-b\hat{H}_B}\rangle$. An alert reader may realize that there is a similar quantity in the context of Wilson loop. If we replace $e^{-a\hat{H}_A}$ and $e^{-b \hat{H}_B}$ by two seperated Wilson loops, then (\ref{correlator}) is the logrithmic of the correlator of two Wilson loops. Expanding (\ref{correlator}) around $(a,b)=(0,0)$, 
\be
T_{A\cup B}(a,b)=\sum_{m.n=1}^\infty \frac{1}{m!n!} T^{(m,n)}_{A\cup B} a^m b^n,
\ee 
the coefficient before $a^m b^n$ is the connected correlation function (\ref{deftmn}). The summation is from $(m,n)=(1,1)$ because $T_{A\cup B}(a,0)=T_{A\cup B}(0,b)=0$. More explicitly, we have 
\be
T_{A\cup B}^{(m,n)}=\frac{\partial^{m+n}T_{A\cup B}(a,b)}{\partial a^m \partial b^n}\big|_{a,b=0},
\ee 
the first few orders are \footnote{Note in our convention, there is an additional minus sign for odd $m+n$ constrast to usual connected correlation functions.}
\bea
T^{(1,1)}_{A\cup B}&=&\langle \hat{H}_A\hat{H}_B\rangle-\langle \hat{H}_A\rangle\langle \hat{H}_B\rangle,\label{t11h}\\
T^{(2,1)}_{A\cup B}&=&-\langle \hat{H}_A^2\hat{H}_B\rangle+\langle \hat{H}_A^2\rangle\langle \hat{H}_B\rangle+2\langle \hat{H}_A\hat{H}_B\rangle\langle \hat{H}_A\rangle-2\langle \hat{H}_A\rangle^2\langle \hat{H}_B\rangle,\label{t21h}\\
T^{(3,1)}_{A\cup B}&=&\langle \hat{H}_A^3\hat{H}_B\rangle-\langle \hat{H}_A^3\rangle\langle \hat{H}_B\rangle-3\langle \hat{H}_A^2\hat{H}_B\rangle\langle \hat{H}_A\rangle-3\langle \hat{H}_A^2\rangle\langle \hat{H}_A\hat{H}_B\rangle+6\langle \hat{H}_A^2\rangle\langle \hat{H}_A\rangle \langle \hat{H}_B\rangle\nn\\
&&+6\langle \hat{H}_A\hat{H}_B\rangle\langle \hat{H}_A\rangle^2-6\langle \hat{H}_A\rangle^3\langle \hat{H}_B\rangle,\label{t31h}\\
T^{(2,2)}_{A\cup B}&=&\langle \hat{H}_A^2\hat{H}_B^2\rangle-2\langle \hat{H}_A^2\hat{H}_B\rangle\langle \hat{H}_B\rangle-2\langle \hat{H}_B^2\hat{H}_A\rangle\langle \hat{H}_A\rangle-2\langle \hat{H}_A\hat{H}_B\rangle^2-\langle \hat{H}_A^2\rangle\langle \hat{H}_B^2\rangle\nn\\&&+2\langle \hat{H}_A^2\rangle\langle \hat{H}_B\rangle^2+2\langle \hat{H}_B^2\rangle\langle \hat{H}_A\rangle^2+8\langle \hat{H}_A\hat{H}_B\rangle\langle \hat{H}_A\rangle\langle \hat{H}_B\rangle-6\langle \hat{H}_A\rangle^2\langle \hat{H}_B\rangle^2.\label{t22h}
\eea 
We don't present the correlators for $m<n$ since the defintion is symmetric under the exchange of $A$ and $B$. Generator (\ref{correlator}) is not easy to evaluate in general.  We will focus on two dimensional massless free scalar theory in the following section and then extend it to general conformal field theories. The advantage of two dimensional conformal system is that the modular Hamiltonian of a single region is well known. 
\section{Two dimensional massless free  scalar}
To get familiar with the concept $T_{A\cup B}(a,b)$, we will compute the generator (\ref{correlator}) in two dimensional massless free scalar system.  The system is in vacuum with Lagrangian 
\be
\mathcal{L}=\frac{1}{2}\partial_{\mu}\phi\partial^{\mu}\phi.
\ee
We use $t,z$ to denote spacetime coordinates
\be
x^{\mu}=(t,z).
\ee 
Region $A$ ($B$) is an interval with radius $R_A(R_B)$ whose center is  at $z_A(z_B)$, 
\bea
A&=&\{(0,z)|x_2\le z\le x_1\},\label{regionA}\\
B&=&\{(0,z)|x_4\le z\le x_3\},\label{regionB}
\eea 
where the end points of the intervals are 
\be
x_1=z_A+R_A,\ x_2=z_A-R_A,\ x_3=z_B+R_B,\ x_4=z_B-R_B.
\ee 
$A$ and $B$ are disjoint, we can assume
\be
x_1>x_2>x_3>x_4.
\ee 
Massless free theory is a conformal field theory, there is a unique cross ratio 
\be
\xi\equiv\frac{x_{12}x_{34}}{x_{13}x_{24}}
\ee 
where 
\be
x_{ij}=x_i-x_j.\label{xijdef}
\ee 
The cross ratio is always between $0$ and $1$, 
\be
0<\xi<1.
\ee 
Another quantity which is related to cross ratio is 
\be
\eta\equiv\frac{x_{12}x_{34}}{x_{14}x_{23}}=\frac{\xi}{1-\xi}.\label{etadef}
\ee 
$\eta$ is between $0$ and $\infty$,
\be
0<\eta<\infty.
\ee 
Modular hamiltonian of region A is 
\be
\hat{H}_A=2\pi \int_{x_2}^{x_1}dz \frac{R_A^2-(z-z_A)^2}{2R_A}T_{tt}(z)\label{hamA}
\ee 
where the stress tensor $T_{tt}$ is 
\be
T_{tt}=\frac{1}{2}[(\partial_t\phi)^2+(\partial_z\phi)^2].\label{stresstt}
\ee 
The integral is evaluated at constant $t=0$ slice. Unfortunately, a direct expansion of the exponential function 
\be
e^{-a\hat{H}_A}
\ee 
leads to divergent terms when two stress tensors $T_{tt}$ collide. However, as we will show below, the divergent terms are canceled in the generator (\ref{correlator}). It would be much easier to work in momentum space for free scalar. In the following subsections, we will first review the quantization of a free scalar in general curved spacetime and then quantize a massless free scalar in an interval. After that, we will discuss the generator (\ref{correlator}). 
\subsection{Massless free scalar field in curved spacetime}
It is useful to review the general framework \cite{DeWitt:1975ys} of a free scalar field in curved spacetime since we will quantize a free scalar field in a subregion of Minkowski spacetime. We just list several key points of this framework. The spacetime we will consider is 
\be
ds^2=-N^2dt^2+G_{ij}dx^idx^j,\label{curvedmetric}
\ee 
where $N$ is called lapse function and $G_{ij}$ is the $d-1$ dimensinoal reduced metric on the hypersurface of constant time $t=t_0$. The Klein-Gordon inner product of two field configurations is 
\be
(\phi_1,\phi_2)_{KG}=-i\int_{\Sigma}d^{d-1}\vec{x}\sqrt{G}n^{\mu}(\phi_1^*\partial_{\mu}\phi_2-\phi_2\partial_\mu\phi_1^*),
\ee 
where $\Sigma$ is the constant time hypersurface and $n^\mu$ is its unit norm vector. Klein-Gordon equation in spacetime  (\ref{curvedmetric}) is solved by a set of complete set of eigenmodes $f_i(x)$. Eigenmodes can be dedomposed and normalized to satisfy 
\be
(f_i,f_j)_{KG}=-(f_i^*,f_j^*)_{KG}=\delta_{ij},\quad (f_i,f_j^*)_{KG}=(f_i^*,f_j)_{KG}=0.\label{completef}
\ee 
Therefore the scalar field $\phi$ could be decomposed in terms of eigenmodes $f_i$ by 
\be
\phi(x)=\sum_i(a_i f_i+a_i^\dagger f_i^*).\label{decomf}
\ee 
We will assume $f_i$'s are positive frequency modes and $f^*_i$'s are negative frequency modes, then coefficients $a_i$ and $a_i^\dagger$ are annihilation and creation operators, respectively. It is easy to show 
\be
[a_i,a_j]=[a_i^\dagger,a_j^\dagger]=0,\quad [a_i,a_j^\dagger]=\delta_{ij}.\label{annia}
\ee 
Therefore, one can define vacuum $|0_{f}\rangle$ as 
\be
a_i|0_{f}\rangle=0,\quad \forall i. 
\ee 
The subscript $f$ shows that the choice of positive frequency modes $f_i$'s determines the vacuum. However, there is no unique choice of positive frequency modes in general. Suppose another complete set of positive frequency modes $g_I$ satisfy 
\be
(g_I,g_J)_{KG}=-(g_I^*,g_J^*)_{KG}=\delta_{IJ},\quad (g_I,g_J^*)_{KG}=(g_I^*,g_J)_{KG}=0.\label{completeg}
\ee
Then the scalar field $\phi$ can also be decomposed in terms of eigenmodes $g_I$ by 
\be
\phi(x)=\sum_I( b_I g_I+b_I^\dagger g_I^*),\label{decomg}
\ee 
where $b_I$ and $b_I^\dagger$ are annihilation and creation operators, respectively. The commutation relations are 
\be
[b_I,b_J]=[b_I^\dagger,b_J^\dagger]=0,\quad [b_I,b_J^\dagger]=\delta_{IJ}.\label{annib}
\ee 
Therefore, one can define vacuum $|0_g\rangle$ by
\be
b_I|0_g\rangle=0,\quad \forall I. 
\ee 
Vacuum $|0_f\rangle$ and $|0_g\rangle$ are not equivalent in general. 
Since $f_i$ eigenmodes are complete, $g_I$ and $g_I^*$ can be decomposed in terms of $f_i$ and $f_i^*$, 
\be
g_I=\sum_i (\alpha_{Ii}f_i+\beta_{Ii}f_i^*),\label{gintermsoff}
\ee 
where $\alpha_{Ii}$ and $\beta_{Ii}$ are called Bogoliubov coefficients. The inverse of  (\ref{gintermsoff}) is 
\be
f_i=\sum_I (\alpha^*_{Ii}g_I-\beta_{Ii}g_I^*)
\ee 
since $g_I$'s are also complete. Then the annihilation and creation operators $a_i,a_i^\dagger,b_I,b_I^\dagger$ are related by Bogoliubov transformation
\bea
a_i&=&\sum_{I}(\alpha_{Ii}b_I+\beta^*_{Ii}b_I^\dagger),\\
a_i^\dagger&=&\sum_{I}(\alpha^*_{Ii}b_I^\dagger+\beta_{Ii}b_I).
\eea 
Inversely, 
\bea
b_I&=&\sum_i(\alpha_{Ii}^*a_i-\beta_{Ii}^*a_i^\dagger),\\
b_I^\dagger&=&\sum_i (\alpha_{Ii}a_i^\dagger-\beta_{Ii}a_i).
\eea 
Therefore, commutation relations (\ref{annia}) and (\ref{annib}) are equivalent to the following consistency relations of Bogoliubov coefficients 
\bea
&&\sum_I\alpha_{Ii}\beta_{Ij}^*-\beta_{Ii}^*\alpha_{Ij}=0,\label{cons1}\\
&&\sum_I\alpha_{Ii}\alpha^*_{Ij}-\beta^*_{Ii}\beta_{Ij}=\delta_{ij}.\label{cons2}
\eea\bea
&&\sum_i \alpha_{Ii}\beta_{Ji}-\beta_{Ii}\alpha_{Ji}=0,\label{cons3}\\
&&\sum_i\alpha_{Ii}^*\alpha_{Ji}-\beta_{Ii}^*\beta_{Ji}=\delta_{IJ}.\label{cons4}
\eea 
$\alpha_{Ii}$ and $\beta_{Ii}$ can be treated as elements of matrices $\bm{\alpha}$ and $\bm{\beta}$, respectively. Then the consistency relations (\ref{cons1}) to (\ref{cons4}) are
\bea
\bm{\alpha}^T\bm{\beta}^*&=&\bm{\beta}^\dagger \bm{\alpha},\quad 
\bm{\alpha}^T\bm{\alpha}^*=1+\bm{\beta}^\dagger \bm{\beta},\label{cons2p}
\eea
and
\bea 
\bm{\alpha}\bm{\beta}^T&=&\bm{\beta}\bm{ \alpha}^T,\quad
\bm{\alpha}\bm{\alpha}^\dagger=1+\bm{\beta}\bm{\beta}^\dagger.\label{cons4p}
\eea  
Finally, before we study massless free scalar field in an interval, we list several results of two dimensional massless free scalar in Minkowski spacetime
\be
ds^2=-dt^2+dz^2. 
\ee 
The field can be decomposed into right moving and left moving modes 
\be
\phi=\phi_R(t-z)+\phi_L(t+z), 
\ee 
where right moving and left moving modes are decouple. We will just consider right moving modes. Solving Klein-Gordon equation, a complete set of eigenmodes is 
\be
f_\omega=N_\omega e^{-i\omega(t-z)},
\ee  
where $\omega$ is frequency which is positive $\omega>0$. 
By choosing $t=0$ time slices and requiring the standard Klein-Gordon bracket relations (\ref{completef}), the normalization constant $N_\omega$ is 
\be
N_\omega=\frac{1}{\sqrt{4\pi \omega}}.
\ee
The scalar field can be decomposed into linear combination of $f_\omega$ and $f_\omega^*$, 
\be
\phi=\sum_{\omega}a_{\omega}f_{\omega}+a^\dagger_{\omega}f^*_{\omega}.
\ee 
The annihilation and creation operators satisfy the commutation relation 
\bea
[a_\omega,a_\omega']=[a_\omega^\dagger,a_{\omega'}^\dagger]=0,\ [a_\omega,a_{\omega'}^\dagger]=\delta(\omega-\omega').
\eea  
Minkowski vacuum $|0_M\rangle$ is annihilated by $a_\omega$,
\be
a_\omega|0_M\rangle=0,\quad \forall\ \omega>0.
\ee 
\subsection{Massless free scalar field in region $A$}
Now we can study a massless free scalar field in region $A$ (\ref{regionA}). The method is similar to quantize a free scalar in Rindler spacetime\cite{Crispino:2007eb}.  The interval  is located at $t=0$ time slice with center position $z_A$ and interval length $2R_A$. The coordinate transformation from Minkowski spacetime to region $A$ is 
\be
t-(z-z_A)=R_A\tanh\frac{\tau-u}{2R_A},\quad 
t+(z-z_A)=R_A\tanh\frac{\tau+u}{2R_A}, \label{tminusz}
\ee 
where $-\infty<\tau, u<\infty $ are new coordinates of region $A$. They cover the causal development of $A$. To see this point, we observe that 
\bea
&\tau\to\pm\infty,\ (t,z)\to (\pm R_A,z_A),\\
&u\to \pm\infty,\ (t,z)\to (0,z_A\pm R_A).
\eea
After the coordinate transformation, the metric becomes 
\be
ds^2=(\cosh\frac{u}{R_A}+\cosh\frac{\tau}{R_A})^{-2}(-d\tau^2+du^2)\label{conformalflat}
\ee 
which is conformally flat. Now we will quantize the massless free boson field in spacetime (\ref{conformalflat}). The Klein-Gordon equation can be solved, the field is still decomposed into right moving and left moving modes. Again, we will just consider right moving modes. A complete set of positive frequency modes is 
\be
g_v=N_v e^{-iv(\tau-u)},
\ee
where $v$ is a positive frequency corresponding to time $\tau$, $v>0$. The normalization constant is still 
\be
N_v=\frac{1}{\sqrt{4\pi v}}. 
\ee 
Therefore, field in region $A$ could be written as 
\be
\phi=\sum_v (b_v g_v+b_v^\dagger g_v^*).
\ee 
The annihilation and creation operators $b_v,b_v^\dagger$ satisfy the commutation relation 
\be
[b_v,b_{v'}]=[b_v^\dagger,b_{v'}^\dagger]=0,\quad\ [b_v,b_{v'}^\dagger]=\delta(v-v').
\ee 
Vacuum in region $A$ is annihilated by $b_v$ 
\be
b_v|0_A\rangle=0,\quad\forall\ v>0.
\ee 
Since field $\phi$ in region $A$ can also be expanded in  terms of Minkowski modes $f_\omega$ given in previous section, the Bogoliubov transformation between $f_\omega$ and $g_v$ modes is 
\be
g_v=\sum_\omega (\alpha_{v\omega}f_\omega+\beta_{v\omega}f_\omega^*).\label{ftog}
\ee 
Since $g_v$'s are not complete in Minkowski spacetime, there is no inverse of transformation (\ref{ftog}). The Bogoliubov coefficients are 
\bea
\alpha_{v\omega}&=&\frac{1}{2\pi}\sqrt{\frac{\omega}{v}}R_A e^{-i\omega z_A}\int_{-1}^1 ds e^{i\omega R_A s}(\frac{1+s}{1-s})^{-ivR_A},\label{boga}\\
\beta_{v\omega}&=&\frac{1}{2\pi}\sqrt{\frac{\omega}{v}}R_A e^{i\omega z_A}\int_{-1}^1 ds e^{-i\omega R_A s}(\frac{1+s}{1-s})^{-ivR_A}.\label{bogb}
\eea 
We checked the consistency conditions (\ref{cons4p}) using Bogoliubov coefficients above. Note (\ref{cons2p}) are not satisfied since $g_v$'s are not complete eigenmodes in Minkowski spacetime. As we have reviewed in previous subsection, annihilation and creation operators in region $A$ are related to those in Minkowski spacetime by Bogoliubov transformation 
\be
b_v=\sum_\omega (\alpha_{v\omega}^* a_\omega-\beta_{v\omega}^* a_\omega^\dagger),\quad b_v^\dagger=\sum_\omega (\alpha_{v\omega} a_\omega^\dagger-\beta_{v\omega} a_\omega)
\ee 
The stress tensor (\ref{stresstt}) in region $A$ can also be casted into right moving and left moving part. Focusing on right moving part, we find the modular Hamiltonian (\ref{hamA}) to be
\be
\hat{H}_A=2\pi R_A \sum_v v b_v^\dagger b_v+\text{const.}\label{ModuA}
\ee 
The form (\ref{ModuA}) is very similar to the Hamiltonian of a free scalar in Minkowski spacetime. The constant term can be fixed by  normalization condition 
\be
1=\text{tr}_A \rho_A=\text{tr}e^{-\hat{H}_A}.
\ee 
Massless free scalar field in region $B$ (\ref{regionB})  is similar, the coordinate transformation from Minkowski spacetime to region $B$ is 
\be
t-(z-z_B)=R_B\tanh\frac{\tau-u}{2R_B},\quad 
t+(z-z_B)=R_B\tanh\frac{\tau+u}{2R_B}. \label{tminuszB}
\ee
Quantizing massless free scalar field, we use $\tilde{v}$ to denote the frequency in region $B$, then the annihilation and creation operators in region $B$ are 
\be
b_{\tilde{v}}=\sum_\omega (\alpha_{\tilde{v}\omega}^* a_\omega-\beta_{\tilde{v}\omega}^* a_\omega^\dagger),\quad b_{\tilde{v}}^\dagger=\sum_\omega (\alpha_{\tilde{v}\omega} a_\omega^\dagger-\beta_{\tilde{v}\omega} a_\omega),
\ee 
where Bogoliubov coefficients are 
\bea
\alpha_{\tilde{v}\omega}&=&\frac{1}{2\pi}\sqrt{\frac{\omega}{\tilde{v}}}R_B e^{-i\omega z_B}\int_{-1}^1 ds e^{i\omega R_B s}(\frac{1+s}{1-s})^{-i\tilde{v}R_B},\label{bogat}\\
\beta_{\tilde{v}\omega}&=&\frac{1}{2\pi}\sqrt{\frac{\omega}{\tilde{v}}}R_B e^{i\omega z_B}\int_{-1}^1 ds e^{-i\omega R_B s}(\frac{1+s}{1-s})^{-i\tilde{v}R_B}.\label{bogbt}
\eea 
Modular Hamiltonian in region $B$ is
\be
\hat{H}_B=2\pi R_B \sum_{\tilde{v}}\tilde{ v} b_{\tilde{v}}^\dagger b_{\tilde{v}}+\text{const.}\label{ModuB}
\ee 
\subsection{Expectation value of an exponential operator}
From the definition of $T_{A\cup B}(a,b)$ and the modular Hamiltonian in momentum space (\ref{ModuA}) and (\ref{ModuB}), the relevant quantity is the expectation value of an exponential operator 
\be
e^{z\sum_I x_I b_I^\dagger b_I},
\ee 
where we can set $x_I$ to be free real function of quantum number $I$ at this moment. For simplicity, we will assume quantum number $I$ to be discrete , the dimension of Hilbert space is finite, we denote the dimension to be $M$. The annihilation and creation operators in the original Minkowski spacetime are also labeled as discrete quantum number $i$ whose dimension is $N$. The result can be easily extended to continuous limit. To introduce the final result, we will clarify some notations at first. We will define two vectors 
$\vec{A}$ and $\vec{B}$ as 
\be
\vec{A}=\left(\begin{array}{c}a_1\\\vdots\\a_N\\a_1^\dagger\\\vdots \\a_N^\dagger\end{array}\right),\quad \vec{B}=\left(\begin{array}{c}b_1\\\vdots\\b_M\\b_1^\dagger\\\vdots\\b_M^\dagger\end{array}\right).
\ee 
Therefore, the commutation relations (\ref{annia}) and (\ref{annib}) are 
\be
[\vec{A}_i,\vec{A}^\dagger_j]=\bm{K}_{ij},\quad [\vec{B}_I,\vec{B}^\dagger_J]=\bm{k}_{IJ}
\ee  
where $\bm{K}$ and $\bm{k}$ are $2N\times 2N$ and $2M\times 2M$ matrices 
\be
\bm{K}=\left(\begin{array}{cc}\bm{1}_{N\times N}&0\\0&-\bm{1}_{N\times N}\end{array}\right),\quad \bm{k}=\left(\begin{array}{cc}\bm{1}_{M\times M}&0\\0&-\bm{1}_{M\times M}\end{array}\right).
\ee 
They are Hermitian matrices. We also note that 
\be
\bm{K}^2=\bm{1}_{2N\times 2N},\quad \bm{k}^2=\bm{1}_{2M\times 2M}.
\ee 
Then the operator $\sum_I x_Ib_I^\dagger b_I$ can be written compactly 
\be
\hat{H}\equiv\sum_I x_I b_I^\dagger b_I=\frac{1}{2}\vec{B}^\dagger \bm{\Lambda}\vec{B}-\frac{1}{2}\text{tr}\bm{X}=\frac{1}{2}\vec{A}^\dagger \bm{H}\vec{A}-\frac{1}{2}\text{tr}\bm{X},\label{H}
\ee 
where $\bm{\Lambda}$ is a $2M\times 2M$ diagonal matrix 
\be
\bm{\Lambda}=\left(\begin{array}{cc}\bm{X}&0\\0&\bm{X}\end{array}\right)
\ee
with 
\be
\bm{X}=\text{diag}(x_1,\cdots,x_M).
\ee  
The Bogoliubov transformation from $a$ modes to $b$ modes is 
\be
\vec{B}=\bm{S}\vec{A},
\ee 
where $\bm{S}$ is a $2M\times 2N$ matrix whose elements are Bogoliubov matrices defined previously 
\be
\bm{S}=\left(\begin{array}{cc}\bm{\alpha}^*&-\bm{\beta}^*\\-\bm{\beta}&\bm{\alpha}\end{array}\right).
\ee 
Its Hermitian conjugate is 
\be
\bm{S}^\dagger=\left(\begin{array}{cc}\bm{\alpha}^T&-\bm{\beta}^\dagger\\-\bm{\beta}^T&\bm{\alpha}^\dagger\end{array}\right).
\ee 
They transform $\bm{K}$ to $\bm{k}$ through consistent relations (\ref{cons4p}),
\be
\bm{S}\bm{K}\bm{S}^\dagger=\bm{k}.\label{sksk}
\ee 
The $2N\times 2N$ matrix $\bm{H}$ in (\ref{H}) can be expressed as 
\be
\bm{H}=\bm{S}^\dagger\bm{\Lambda}\bm{S}=\left(\begin{array}{cc}\bm{\alpha}^T\bm{X}\bm{\alpha}^*+\bm{\beta}^\dagger\bm{X}\bm{\beta}&-\bm{\alpha}^T\bm{X}\bm{\beta}^*-\bm{\beta}^\dagger\bm{X}\bm{\alpha}\\-\bm{\beta}^T\bm{X}\bm{\alpha}^*-\bm{\alpha}^\dagger\bm{X}\bm{\beta}&\bm{\alpha}^\dagger\bm{X}\bm{\alpha}+\bm{\beta}^T\bm{X}\bm{\beta}^*\end{array}\right).\label{bmH}
\ee 
$\bm{H}$ is Hermitian 
\be
\bm{H}=\bm{H}^\dagger.
\ee
The commutation relation of $\hat{H}$ and $\vec{A}$ is 
\be
[\hat{H},\vec{A}]=-\bm{K}\bm{H}\vec{A}.\label{comHA}
\ee
By using Baker-Hausdorff formula and (\ref{comHA}), we have 
\be
e^{z\hat{H}}\vec{A}e^{-z\hat{H}}=e^{-z\bm{K}\bm{H}}\vec{A},
\ee 
where 
\be
e^{-z\bm{K}\bm{H}}=1+\bm{K}\bm{S}^\dagger\bm{k}(e^{-z\bm{k}\bm{\Lambda}}-1)\bm{S}.\label{ezkh}
\ee 
We will prove this identity in Appendix A.  Using the definition of $\bm{K},\bm{k},\bm{\Lambda}$ and $\bm{S}$, $e^{-z\bm{K}\bm{H}}$ can be written as 
\be
e^{-z\bm{K}\bm{H}}=\left(\begin{array}{cc}\bm{\Theta}&\bm{\Phi}\\\bm{\tilde{\Phi}}&\bm{\tilde{\Theta}}\end{array}\right),
\ee 
where $\bm{\Theta},\bm{\Phi},\bm{\tilde{\Theta}},\bm{\tilde{\Phi}}$ are $N\times N$ matrices. $\bm{\Theta}$ is 
\be
\bm{\Theta}=\bm{1}+\bm{\alpha}^T\bm{p}\bm{\alpha}^*+\bm{\beta}^\dagger\bm{q}\bm{\beta},
\ee
where $\bm{p}$ and $\bm{q}$ are rank $M$ diagonal matrices 
\be
\bm{p}=e^{-z\bm{X}}-\bm{1},\quad \bm{q}=\bm{1}-e^{z\bm{X}}.
\ee  
The matrices $\bm{\Phi},\bm{\tilde{\Theta}},\bm{\tilde{\Phi}}$ are not relevant to our result below. 
The expectation value of $e^{z\hat{H}}$ is 
\be
\langle0_M|e^{z\hat{H}}|0_M\rangle=\frac{e^{-\frac{z}{2}\text{tr}\bm{X}}}{\sqrt{\det\bm\Theta}}.\label{exp}
\ee 
We will use normal ordering and parameter differentiation method to prove this identity in Appendix B. We can simplify (\ref{exp}) further by noticing the so called matrix determinant lemma \cite{matrix}, 
\be
\det(\bm{Q}+\bm{u}\bm{W}\bm{v}^T)=\det(\bm{W}^{-1}+\bm{v}^T\bm{Q}^{-1}\bm{u})\det \bm{W}\det\bm{Q},
\ee 
where $\bm{Q}$ is an $N\times N$ invertible matrix, $\bm{W}$ is an $M\times M$ invertible matrix, $\bm{u},\bm{v}$ are $N\times M$ matrices. Note $\bm{\Theta}$ is exactly the form of $\bm{Q}+\bm{u}\bm{W}\bm{v}^T$, 
\be
\bm{\Theta}=\bm{1}+(\bm{\alpha}^T,\bm{\beta}^\dagger)\left(\begin{array}{cc}\bm{p}&0\\0&\bm{q}\end{array}\right)\left(\begin{array}{c}\bm{\alpha}^*\\\bm{\beta}\end{array}\right).
\ee 
Therefore 
\bea
\det\bm{\Theta}&=&\det[\bm{1}+\left(\begin{array}{cc}\bm{p}&0\\0&\bm{q}\end{array}\right)\left(\begin{array}{c}\bm{\alpha}^*\\\bm{\beta}\end{array}\right)(\bm{\alpha}^T,\bm{\beta}^\dagger)]\nn\\
&=&\det\left(\begin{array}{cc}\bm{1}+\bm{p}\bm{\alpha}^*\bm{\alpha}^T&\bm{p}\bm{\alpha}^*\bm{\beta}^\dagger\\\bm{q}\bm{\beta}\bm{\alpha}^T&\bm{1}+\bm{q}\bm{\beta}\bm{\beta}^\dagger\end{array}\right)\nn\\
&=&\det(\bm{1}+\bm{p})\det \bm{T},\label{bmTheta}
\eea
where the matrix $\bm{T}$ is 
\be
\bm{T}=\left(\begin{array}{cc}\bm{1}+\bm{q}\bm{\beta}^*\bm{\beta}^T&\bm{q}\bm{\alpha}^*\bm{\beta}^\dagger\\\bm{q}\bm{\beta}\bm{\alpha}^T&\bm{1}+\bm{q}\bm{\beta}\bm{\beta}^\dagger\end{array}\right).\label{bmZ}
\ee 
At the last step of (\ref{bmTheta}), we used the Bogoliubov coefficients consistency conditions (\ref{cons4p}) and 
\be
(\bm{1}+\bm{p})^{-1}\bm{p}=\bm{q}.\ee
Now since 
\be
\det(\bm{1}+\bm{p})=\det e^{-z\bm{X}}=e^{-z\text{tr}\bm{X}}, 
\ee 
we can simplify (\ref{exp}) to be 
\be
\langle 0_M|e^{z\hat{H}}|0_M\rangle=\frac{1}{\sqrt{\det\bm{T}}}.
\ee 
\subsection{Region $A$}
From previous discussion, the expectation value of an exponential  operator of the form $\hat{H}=\sum_I x_I b_I^\dagger b_I$ is determined by a matrix $\bm{T}$ which is defined by (\ref{bmZ}). We denote 
\be
T_A(a)=\log \langle0_M|e^{-a \hat{H}_A}|0_M\rangle=-\frac{1}{2}\log\det\bm{T}_A(a),\label{ZAa}
\ee 
then the matrix $\bm{T}_A(a)$ is 
\be
\bm{T}_A(a)=\left(\begin{array}{cc}\bm{1}+\bm{q}_A(a)\bm{\beta}_A^*\bm{\beta}_A^T&\bm{q}_A(a)\bm{\alpha}_A^*\bm{\beta}_A^\dagger\\\bm{q}_A(a)\bm{\beta}_A\bm{\alpha}_A^T&\bm{1}+\bm{q}_A(a)\bm{\beta}_A\bm{\beta}_A^\dagger\end{array}\right),\label{bmZA}
\ee 
where $\bm{1}$ should be understood as Dirac delta function in the continuous limit,
\be
\bm{1}_{vv'}=\delta(v-v').
\ee 
The diagonal matrix $\bm{q}_A(a)$ has the following elements
\be
( \bm{q}_A(a))_{vv'}=(1-e^{-2\pi a v R_A})\delta(v-v').
\ee 
Using Bogoliubov coefficients  (\ref{boga}) and (\ref{bogb}), we find 
\bea
(\bm{\beta}_A^*\bm{\beta}_A^T)_{vv'}&=&\sum_\omega(\beta_A)^*_{v\omega}(\beta_A)_{v'\omega}=\frac{\delta(v-v')}{e^{2\pi v R_A}-1},\label{bst}\\
(\bm{\beta}_A\bm{\beta}_A^\dagger)_{vv'}&=&\sum_\omega(\beta_A)_{v\omega}(\beta_A)^*_{v'\omega}=\frac{\delta(v-v')}{e^{2\pi v R_A}-1},\label{bda}\\
(\bm{\alpha}^*_A\bm{\beta}_A^\dagger)_{vv'}&=&\sum_{\omega}(\alpha_A)^*_{v\omega}(\beta_A)^*_{v'\omega}=0,\label{asbs}\\
(\bm{\beta}_A\bm{\alpha}_A^T)_{vv'}&=&\sum_{\omega}(\beta_A)_{v\omega}(\alpha_A)^*_{v'\omega}=0.\label{bat}
\eea
The details of the computation is in Appendix C. 
So $\bm{T}_A$ is a diagonal matrix. 
Therefore (\ref{ZAa}) is 
\bea
T_A(a)&=&-\frac{1}{2}\text{tr}\log\bm{T}_A(a)=-\text{tr}\log(1+\frac{1-e^{-2\pi a v R_A}}{e^{2\pi v R_A}-1})\delta(0)\nn\\&=&-\int_0^\infty dv \log\frac{1-e^{-2\pi(1+a)vR_A}}{1-e^{-2\pi v R_A}}\delta(0)\nn\\&=&-\frac{\pi}{12R_A}\frac{a}{1+a}\delta(0),
\eea 
where $\delta(0)$ is from the diagonal element in momentum space, formally it is Dirac delta function evaluated at 0,
\be
\delta(0)=\delta(v-v')|_{v'=v}.
\ee  
We will regularize it by matching $Z_A(a)$ to R\'enyi entropy of region $A$. This provides a consistent check of our result.  We already know the reduced density matrix $\rho_A$ is 
\be
\rho_A=\rho_0 e^{-\hat{H}_A},
\ee 
where $\rho_0$ can be fixed by the normalization of $\rho_A$
\be
1=\rho_0 \text{tr}_A e^{-\hat{H}_A}
\ee 
In momentum space, the right hand side is easy to calculate, 
therefore 
\be
\rho_0=\prod_v(1-e^{-2\pi v R_A})\delta(0).
\ee 
Then 
\bea
S_A^{(n)}&=&\frac{\log\text{tr}_A\rho_A^n}{1-n}=\frac{\log \langle 0_M|\rho_A^{n-1}|0_M\rangle}{1-n}\nn\\&=&2\times(-\sum_v \log(1-e^{-2\pi v R_A})\delta(0)+\frac{\log \text{tr}_A\rho_A e^{-(n-1)H_A}}{1-n})\nn\\&=&2\times(\frac{\pi}{12R_A}\delta(0)+\frac{\log\langle0_M|e^{-(n-1)H_A}|0_M\rangle}{1-n})\nn\\&=&2\times(\frac{\pi}{12R_A}\delta(0)+\frac{1}{1-n}T_A(n-1))\nn\\&=&\frac{\pi}{6R_A}(1+\frac{1}{n})\delta(0).\label{Renyi1}
\eea 
A factor $2$ is inserted at the intermediate step to count the left moving and right moving modes. 
On the other hand, a general two dimensional conformal field theory with central charge $c$  has R\'enyi entropy 
\be
S_A^{(n)}=\frac{c}{6}(1+\frac{1}{n})\log\frac{2R_A}{\epsilon}\label{Renyi2}
\ee 
for one inverval, where $\epsilon$ is a UV cutoff. Comparing (\ref{Renyi1}) with (\ref{Renyi2}) and noticing that the central charge of free boson is 1, we have the following regularization rule
\be
\delta(0)\to \frac{R_A}{\pi}\log\frac{2R_A}{\epsilon}.
\ee 
\subsection{Region $A$ and $B$}
We are interested in the expectation value of the following operator 
\be
e^{-a\hat{H}_A-b\hat{H}_B}=e^{-a \sum_v v b_v^\dagger b_v-b \sum_{\tilde{v}}\tilde{v}b^\dagger_{\tilde{v}}b_{\tilde{v}}}.
\ee 
From previous general discussion, we find that 
\be
\log \langle 0_M|e^{-a \hat{H}_A-b\hat{H}_B}|0_M\rangle=-\frac{1}{2}\log \det \bm{T}_{A\cup B}(a,b),
\ee 
where
\be
\bm{T}_{A\cup  B}(a,b)=\left(                 
\begin{array}{cccc}   
M_{AA}& M_{AB} & C_{AA}&C_{AB}\\  
M_{BA} & M_{BB} & C_{BA}&C_{BB}\\  
D_{AA}&D_{AB}&N_{AA}&N_{AB}\\
D_{BA}&D_{BB}&N_{BA}&N_{BB}
\end{array}
\right) .\label{TABaboriginal}
\ee
The matrix $M,N,C,D$ are 
\begin{align}
&M_{AA}=\bm{1}+\bm{q}_A(a)\bm{\beta}^*_A\bm{\beta}^T_A,&& M_{BB}=\bm{1}+\bm{q}_B(b)\bm{\beta}^*_B\bm{\beta}^T_B,\\
&M_{AB}=\bm{q}_A(a)\bm{\beta}^*_A\bm{\beta}^T_B,&& M_{BA}=\bm{q}_B(b)\bm{\beta}^*_B\bm{\beta}^T_A,\\
&N_{AA}=\bm{1}+\bm{q}_A(a)\bm{\beta}_A\bm{\beta}^\dagger_A,&&N_{BB}=\bm{1}+\bm{q}_B(b)\bm{\beta}_B\bm{\beta}_B^\dagger,\\
&N_{AB}=\bm{q}_A(a)\bm{\beta}_A\bm{\beta}^\dagger_B,&&N_{BA}=\bm{q}_B(b)\bm{\beta}_B\bm{\beta}^\dagger_A,\\
&C_{AA}=\bm{q}_A(a)\bm{\alpha}^*_A\bm{\beta}^\dagger_A,&&C_{BB}=\bm{q}_B(b)\bm{\alpha}^*_B\bm{\beta}^\dagger_B,\\
&C_{AB}=\bm{q}_{A}(a)\bm{\alpha}^*_A\bm{\beta}^\dagger_B,&&C_{BA}=\bm{q}_B(b)\bm{\alpha}^*_B\bm{\beta}^\dagger_A,\\
&D_{AA}=\bm{q}_A(a)\bm{\beta}_A\bm{\alpha}^T_A,&&D_{BB}=\bm{q}_B(b)\bm{\beta}_B\bm{\beta}^T_B,\\
&D_{AB}=\bm{q}_A(a)\bm{\beta}_A\bm{\alpha}^T_B,&&D_{BA}=\bm{q}_B(b)\bm{\beta}_B\bm{\alpha}^T_A.
\end{align}
Some of the matrices are already discussed in previous section, 
\be
\bm{T}_A(a)=\left(\begin{array}{cc}
M_{AA}&C_{AA}\\
D_{AA}&N_{AA}
\end{array}
\right),\quad \bm{T}_B(b)=\left(\begin{array}{cc}
M_{BB}&C_{BB}\\
D_{BB}&N_{BB}
\end{array}
\right),
\ee 
where 
\be
C_{AA}=D_{AA}=C_{BB}=D_{BB}
\ee 
and
\bea
(M_{AA})_{vv'}&=&(N_{AA})_{vv'}=\frac{1-e^{-2\pi v R_A(1+a)}}{1-e^{-2\pi v R_A}}\delta(v-v'),\\
(M_{BB})_{\tilde{v}\tilde{v}'}&=&(N_{BB})_{\tilde{v}\tilde{v}'}=\frac{1-e^{-2\pi \tilde{v}R_B(1+b)}}{1-e^{-2\pi \tilde{v}R_B}}\delta(\tilde{v}-\tilde{v}').
\eea 
Matrices $\bm{q}_A$ and $\bm{q}_B$ are
\bea
(\bm{q}_A(a))_{vv'}=(1-e^{-2\pi v R_A a})\delta(v-v'),\quad (\bm{q}_B(b))_{\tilde{v}\tilde{v}'}=(1-e^{-2\pi \tilde{v} R_B b})\delta(\tilde{v}-\tilde{v}').
\eea 
Bogoliubov matrices which connect region A and region B are 
\begin{align}
&(\bm{\beta}_A^*\bm{\beta}_B^T)_{v\tilde{v}}=G(ivR_A,-i\tilde{v}R_B),\quad&&(\bm{\beta}_A\bm{\beta}_B^\dagger)_{v\tilde{v}}=G(-ivR_A,i\tilde{v}R_B),\\
&(\bm{\beta}_A\bm{\alpha}^T_B)_{v\tilde{v}}=G(-ivR_A,-i\tilde{v}R_B),\quad&&(\bm{\alpha}_A^*\bm{\beta}_B^\dagger)_{v\tilde{v}}=G(ivR_A,i\tilde{v}R_B),\\
&(\bm{\beta}_B^*\bm{\beta}_A^T)_{\tilde{v}v}=G(-ivR_A,i\tilde{v}R_B),\quad&&(\bm{\beta}_B\bm{\beta}^\dagger_A)_{\tilde{v}v}=G(ivR_A,-i\tilde{v}R_B),\\
&(\bm{\beta}_B\bm{\alpha}^T_A)_{\tilde{v}v}=G(-ivR_A,-i\tilde{v}R_B),\quad&&(\bm{\alpha}_B^*\bm{\beta}^\dagger_A)_{\tilde{v}v}=G(ivR_A,i\tilde{v}R_B),
\end{align}where the function $G(x,y)$ is 
\be
G(x,y)=-\frac{\sqrt{R_AR_B}}{4\pi^2\sqrt{|xy|}}B(1+x,1-x)B(1+y,1-y)\eta x_{23}^{-x}x_{14}^{-y}x_{13}^{x+y}{}_2F_1(1+x,1+y,2,-\eta).
\ee 
Please find the details of the Bogoliubov matrices in Appendix C. Since $T_{A\cup B}(a,b)$ only depends on the determinant of $\bm{T}_{A\cup B}(a,b)$, we can write $\bm{T}_{A\cup B}(a,b)$ as \footnote{ If an $n\times n$ matrix $\mathbf{M}$ is formed from an $n\times n$ matrix $\mathbf{M}'$ by interchanging two rows or two columns of $\mathbf{M}'$, then 
\be
\det \mathbf{M}=-\det\mathbf{M}'.
\ee See chapter 13 of \cite{matrix}.  The matrix (\ref{TABabexhange}) is obtained from (\ref{TABaboriginal}) by interchanging two rows (or two columns) even times, therefore the determinant is invariant. }
\be
\bm{T}_{A\cup B}(a,b)=\left(                 
\begin{array}{cccc}   
M_{AA}& C_{AA} & M_{AB}&C_{AB}\\  
D_{AA} & N_{AA} & D_{AB}&N_{AB}\\  
M_{BA}&C_{BA}&M_{BB}&C_{BB}\\
D_{BA}&N_{BA}&D_{BB}&N_{BB}
\end{array}
\right) =\left(                 
\begin{array}{cccc}   
M_{AA}& 0 & M_{AB}&C_{AB}\\  
0 & N_{AA} & D_{AB}&N_{AB}\\  
M_{BA}&C_{BA}&M_{BB}&0\\
D_{BA}&N_{BA}&0&N_{BB}
\end{array}
\right) \label{TABabexhange}
\ee 
without changing the value of $T_{A\cup B}(a,b)$. Therefore 

\bea
T_{A\cup B}(a,b)&=&-\frac{1}{2}\log\det \bm{T}_{A\cup B}(a,b)+\frac{1}{2}\log\det \bm{T}_A(a)+\frac{1}{2}\log\det\bm{T}_B(b)\nn\\&=&-\frac{1}{2}\text{tr}\log[\left(\begin{array}{cc}\bm{T}_A(a)^{-1}&0\\0&\bm{T}_B(b)^{-1}\end{array}\right)\left(\begin{array}{cc}\bm{T}_{A}(a)&U\\V&\bm{T}_{B}(b)\end{array}\right)]\nn\\&=&-\frac{1}{2}\text{tr}\log[\bm{1}+\left(\begin{array}{cc}0&\bm{T}_{A}(a)^{-1}U\\\bm{T}_B(b)^{-1}V&0\end{array}\right)]\nn\\&=&-\frac{1}{2}\text{tr}\log[\bm{1}-\bm{T}_A(a)^{-1}U\bm{T}_B(b)^{-1}V],\label{TrAB}
\eea
where 
\be
U=\left(\begin{array}{cc}M_{AB}&C_{AB}\\D_{AB}&N_{AB}\end{array}\right),\quad V=\left(\begin{array}{cc}M_{BA}&C_{BA}\\D_{BA}&N_{BA}\end{array}\right).
\ee 
The matrix $\bm{T}_A(a)^{-1}U\bm{T}_B(b)^{-1}V$ is 
\be
\bm{T}_A(a)^{-1}U\bm{T}_B(b)^{-1}V=\left(\begin{array}{cc}\mathcal{A}&\mathcal{C}\\\mathcal{D}&\mathcal{B}\end{array}\right)
\ee
where 
\bea
\mathcal{A}&=&M_{AA}^{-1}M_{AB}M_{BB}^{-1}M_{BA}+M_{AA}^{-1}C_{AB}N_{BB}^{-1}D_{BA},\\
\mathcal{B}&=&N_{AA}^{-1}D_{AB}M_{BB}^{-1}C_{BA}+N_{AA}^{-1}N_{AB}N_{BB}^{-1}N_{BA},\\
\mathcal{C}&=&M_{AA}^{-1}M_{AB}M_{BB}^{-1}C_{BA}+M_{AA}^{-1}C_{AB}N_{BB}^{-1}N_{BA},\\
\mathcal{D}&=&N_{AA}^{-1}D_{AB}M_{BB}^{-1}M_{BA}+N_{AA}^{-1}N_{AB}N_{BB}^{-1}D_{BA}.
\eea
Since $M_{AA},M_{BB},N_{AA},N_{BB}$ are diagonal matrices, we could easy obtain their inverse
\bea
(M_{AA})^{-1}_{vv'}&=&(N_{AA})^{-1}_{vv'}=\frac{1-e^{-2\pi v R_A}}{1-e^{-2\pi v R_A(1+a)}}\delta(v-v'),\\
(M_{BB})^{-1}_{\tilde{v}\tilde{v}'}&=&(N_{BB})^{-1}_{\tilde{v}\tilde{v}'}=\frac{1-e^{-2\pi \tilde{v}R_B}}{1-e^{-2\pi \tilde{v}R_B(1+b)}}\delta(\tilde{v}-\tilde{v}').
\eea 
Therefore, $\mathcal{A},\mathcal{B},\mathcal{C}$ and $\mathcal{D}$ are 
\bea
\hspace{-50pt}\mathcal{A}_{vv'}\hspace{-10pt}&=&\hspace{-10pt}\frac{R_A\eta^2}{4}\hspace{-5pt}\int_0^\infty \hspace{-10pt}dy \frac{\sqrt{xx'}y\sinh\pi a x\ \sinh\pi b y}{\sinh \pi x'\ \sinh \pi y\ \sinh\pi(1+a)x\ \sinh\pi(1+b)y}(\frac{x_{13}}{x_{23}})^{i(x-x')}\mathcal{F}(x,x',y),\\
\hspace{-50pt}\mathcal{B}_{vv'}\hspace{-10pt}&=&\hspace{-10pt}\frac{R_A\eta^2}{4}\hspace{-5pt}\int_0^\infty \hspace{-10pt}dy \frac{\sqrt{xx'}y\sinh\pi a x\ \sinh\pi b y}{\sinh \pi x'\ \sinh \pi y\ \sinh\pi(1+a)x\ \sinh\pi(1+b)y}(\frac{x_{13}}{x_{23}})^{-i(x-x')}\mathcal{F}(x',x,y),\\
\hspace{-50pt}\mathcal{C}_{vv'}\hspace{-10pt}&=&\hspace{-10pt}\frac{R_A\eta^2}{4}\hspace{-5pt}\int_0^\infty\hspace{-10pt} dy \frac{\sqrt{xx'}y\sinh\pi a x\ \sinh\pi b y}{\sinh \pi x'\ \sinh \pi y\ \sinh\pi(1+a)x\ \sinh\pi(1+b)y}(\frac{x_{13}}{x_{23}})^{i(x+x')}\mathcal{F}(x,-x',y),\\
\hspace{-50pt}\mathcal{D}_{vv'}\hspace{-10pt}&=&\hspace{-10pt}\frac{R_A\eta^2}{4}\hspace{-5pt}\int_0^\infty\hspace{-10pt} dy \frac{\sqrt{xx'}y\sinh\pi a x\ \sinh\pi b y}{\sinh \pi x'\ \sinh \pi y\ \sinh\pi(1+a)x\ \sinh\pi(1+b)y}(\frac{x_{13}}{x_{23}})^{-i(x+x')}\mathcal{F}(-x,x',y).
\eea with 
\bea
\mathcal{F}(x,x',y)&=&{}_2F_1(1+ix,1-iy,2,-\eta)\ {}_2F_1(1-ix',1+iy,2,-\eta)\nn\\&&+{}_2F_1(1+ix,1+iy,2,-\eta)\ {}_2F_1(1-ix',1-iy,2,-\eta)
\eea
and 
\be
x=v R_A,\quad x'=v' R_A.
\ee
$\mathcal{F}$ and its complex conjugate obey
\be
\mathcal{F}^*(x,x',y)=\mathcal{F}(x',x,y),\quad \mathcal{F}^*(-x,-x',y)=\mathcal{F}(x,x',y).
\ee
so 
\be
\mathcal{A}=\mathcal{B}^*,\quad \mathcal{C}=\mathcal{D}^*.
\ee
The compact form of (\ref{TrAB}) can be evaluated as a series expansion 
\bea
T_{A\cup B}(a,b)&=&\sum_{n=1}^\infty T_n(a,b)=\frac{1}{2}\sum_{n=1}^\infty \frac{1}{n}\text{tr}\bm{T}_n,\label{TrABexp1}
\eea where we define
\bea
T_n(a,b)=\frac{1}{2n}\text{tr}\bm{T}_n,\quad \bm{T}_n=\left(\begin{array}{cc}\mathcal{A}&\mathcal{C}\\\mathcal{D}&\mathcal{B}\end{array}\right)^n\label{TrABexp2}.
\eea 
Given the exact result (\ref{TrABexp1})-(\ref{TrABexp2}), we discuss several properties of $T_{A\cup B}(a,b)$ for two dimensional massless free scalar in the following.
\subsubsection{Large distance expansion}
When two regions $A$ and $B$ are far way to each other, $\eta\ll1$, then 
the  matrix
\be
{T}_n(a,b)\sim \mathcal{O}(\eta^{2n}).
\ee  Therefore (\ref{TrABexp1}) can be understood as large distance expansion of $T_{A\cup B}(a,b)$. As $n$ increases, the contribution of ${T}_n$ decreases. The leading term is 
\bea
T_1(a,b)&=&\frac{1}{2}\text{tr}(\mathcal{A}+\mathcal{B})\nn\\&=&\frac{\eta^2}{4}\int_0^{\infty}dx\int_0^{\infty}dy\frac{x y\sinh\pi a x\ \sinh\pi b y }{\sinh\pi x\ \sinh\pi y\ \sinh\pi(1+a)x\ \sinh\pi(1+b)y}
\times\mathcal{F}(x,x,y)\nn\\\label{leadingTab}
\eea
with 
\be
\mathcal{F}(x,x,y)=|{}_2F_1(1+ix,1+i y,2,-\eta)|^2+|{}_2F_1(1+ix,1-i y,2,-\eta)|^2.
\ee
We don't find an easy way to write $T_1(a,b)$ as special functions. However, we check numerically that $T_1(a,b)$ is finite for general positive values of $a$ and $b$. $T_1(a,b)$ is symmetric under the exchange of $a$ and $b$ 
\be
T_1(a,b)=T_1(b,a).
\ee
This is because region $A$ and $B$ are symmetric in the definition of $T_{A\cup B}(a,b)$. $T_1(a,b)$
is obviously zero whenever $a=0$ or $b=0$
\be
T_1(a,0)=T_1(0,b)=0
\ee 
which is also expected from the definition of $T_{A\cup B}(a,b)$.

In the large distance limit, the first two terms of $\mathcal{F}(x,x,y)$ are independent of $x$ and $y$ 
\be
\mathcal{F}(x,x,y)=2(1-\eta+\mathcal{O}(\eta^2)),
\ee 
therefore 
\bea
T_1(a,b)&=&\frac{\eta^2}{2}\int_0^{\infty}dx\int_0^{\infty}dy\frac{x y\sinh\pi a x\ \sinh\pi b y }{\sinh\pi x\ \sinh\pi y\ \sinh\pi(1+a)x\ \sinh\pi(1+b)y}(1-\eta+\mathcal{O}(\eta^2))\nn\\
&=&\frac{a(a+2)b(b+2)}{288(1+a)^2(1+b)^2}(\eta^2-\eta^3+\mathcal{O}(\eta^4)).\label{T1ablarge}
\eea 
The first two terms are also the leading two terms of $T_{A\cup B}(a,b)$ since $T_n(a,b)$ is at least $\mathcal{O}(\eta^4)$ for any $n\ge2$. Now we compute $T_2(a,b)$, 
\bea
T_2(a,b)&=&\frac{1}{4}\text{tr}(\mathcal{A}^2+\mathcal{B}^2+2\mathcal{C}\mathcal{D})\nn\\&=&\frac{\eta^4}{32}\int_0^\infty dx\frac{x\sinh\pi a x}{\sinh\pi x\ \sinh\pi (1+a)x}\int_0^\infty dx' \frac{x'\sinh\pi a x'}{\sinh\pi x'\ \sinh\pi (1+a)x'}\nn\\&&\times\int_0^\infty dy\frac{y\sinh\pi b y}{\sinh\pi y\ \sinh\pi (1+b)y}\int_0^\infty dy' \frac{y'\sinh\pi b y'}{\sinh\pi y'\ \sinh\pi (1+b)y'}\nn\\&&\times
\mathcal{F}_2(x,x',y,y'),
\eea where
\be
\mathcal{F}_2(x,x',y,y')= \mathcal{F}(x,x',y)\mathcal{F}(x',x,y')+\mathcal{F}(x,-x',y)\mathcal{F}(-x',x,y').
\ee 
As $T_1(a,b)$, $T_2(a,b)$ is symmetric under the exchange of $a$ and $b$. We also check numerically that $T_2(a,b)$ is finite for general positive values of $a$ and $b$.
\subsubsection{Correlation functions of modular Hamiltonians}
From modular hamiltonians $\hat{H}_A$ and $\hat{H}_B$, one can define correlation functions 
\be
\langle \hat{H}_A^m\hat{H}_B^n\rangle\label{HAmHBn}
\ee 
for any positive integers $m$ and $n$. These correlation functions are divergent in general when $m\ge2$ or $n\ge2$. However, from section 2, we notice that $T_{A\cup B}(a,b)$ is the generator of correlation functions of (\ref{HAmHBn}). More explicitly, it is the generator of connected correlation functions 
\be
\langle \hat{H}_A^m \hat{H}_B^n\rangle_c\equiv T_{A\cup B}^{(m,n)}\label{HABmHABnc}
\ee 
which removes divergent terms. We will discuss the correlation functions (\ref{HABmHABnc}) in two dimensional massless free scalar theory. We first 
define a set of quantities
\be
T_k^{(m,n)}=\frac{\partial^{m+n}}{\partial a^m\partial b^n}T_k(a,b)|_{a=0,b=0}.\quad k\ge 1.
\ee 
Using (\ref{TrABexp2}) and the matrices $\mathcal{A},\mathcal{B},\mathcal{C},\mathcal{D}$, 
it is easy to convince oneself that 
\be
\left\{\begin{aligned}&T_k^{(m,n)}=0,\quad m\le k-1\ \text{or}\ n\le k-1,\\&T_k^{(m,n)}\not=0,\quad m\ge k\ \text{and}\ n\ge k.\end{aligned}\right.\label{tkmn}
\ee
Therefore,
\bea
T_{A\cup B}^{(m,n)}=\sum_{k=1}^\infty T_k^{(m,n)}=\sum_{k=1}^{\text{min}(m,n)}T_k^{(m,n)},\ m\ge1,n\ge1.
\eea 
At the second step, we used (\ref{tkmn}). So to obtain correlation functions (\ref{HABmHABnc}), only  finite number of terms contribute for any fixed $m$ and $n$. We list first few terms below 
\bea
T_{A\cup B}^{(1,1)}&=&T_1^{(1,1)}=\frac{\pi^2\eta^2}{4}\int_0^\infty dx\int_0^\infty dy \frac{x^2y^2}{\sinh^2\pi x\sinh^2\pi y}\mathcal{F}(x,x,y)\nn\\&=&-\frac{1}{6}+\frac{(2+\eta)\log(1+\eta)}{12\eta},\label{tab11}\\
T_{A\cup B}^{(2,1)}&=&T_1^{(2,1)}=-\frac{\pi^3\eta^2}{2}\int_0^\infty dx\int_0^\infty dy \frac{x^3y^2\coth\pi x}{\sinh^2\pi x\sinh^2\pi y}\mathcal{F}(x,x,y)\nn\\&=&\frac{1}{2}-\frac{(2+\eta)\log(1+\eta)}{4\eta},\label{tab21}\\
T_{A\cup }^{(3,1)}&=&T_1^{(3,1)}=\frac{\pi^4\eta^2}{2}\int_0^\infty dx\int_0^\infty dy \frac{x^4y^2(2+\cosh 2\pi x)}{\sinh^4\pi x\sinh^2\pi y}\mathcal{F}(x,x,y)\nn\\&=&-2+\frac{(2+\eta)\log(1+\eta)}{\eta},\label{tab31},\label{tab31}\\
T_{A\cup B}^{(2,2)}&=&T_1^{(2,2)}+T_2^{(2,2)}\nn\\&=&\pi^4\eta^2\int_0^\infty dx\int_0^\infty dy \frac{x^3y^3\coth\pi x\coth\pi y}{\sinh^2\pi x\sinh^2\pi y}\mathcal{F}(x,x,y)\nn\\&&+\frac{\pi^4\eta^4}{8}\int_0^\infty dx\int_0^\infty dx'\int_0^\infty dy\int_0^\infty dy'\frac{x^2x'^2y^2y'^2}{\sinh^2\pi x\sinh^2\pi x'\sinh^2\pi y\sinh^2\pi y'}\mathcal{F}_2(x,x',y,y').\label{tab22}\nn\\
\eea 
At the second line of (\ref{tab11}),(\ref{tab21}) and (\ref{tab31}), we check the integals numerically. We don't find a simple function which is equivalent to the integral (\ref{tab22}). However, in the large distance limit, 
the integral can be done term by term 
\be
T^{(2,2)}_{A\cup B}(0,0)=\frac{\eta^2}{8} - \frac{\eta^3}{8} + \frac{37 \eta^4}{324} - \frac{67 \eta^5}{648} + \frac{
4061 \eta^6}{43200}+\mathcal{O}(\eta^7).\label{larget22}
\ee
Interestingly, 
\be
T_{A\cup B}^{(1,1)}=-\frac{1}{3}T_{A\cup B}^{(2,1)}=\frac{1}{12}T_{A\cup B}^{(3,1)}.
\ee 
We can also compute $T_{A\cup B}^{(m,1)}$ 
\be
T_{A\cup B}^{(m,1)}=T_1^{(m,1)}.
\ee 
We checked numerically that $T_{A\cup B}^{(m,1)}$ is always proportional to $T_{A\cup B}^{(1,1)}$ for any $m\ge 1$. Actually, 
\bea
T_{A\cup B}(a)&\equiv& \frac{\partial}{\partial b}T^r_{A\cup B}(a,b)|_{b=0}\nn\\
&=&\frac{\partial}{\partial b} T_1(a,b)|_{b=0}\nn\\
&=&\frac{\pi \eta^2}{4}\int_0^\infty dx \int_0^\infty dy \frac{x y^2\sinh\pi a x}{\sinh\pi x\sinh\pi(1+a)x\sinh^2\pi y}\mathcal{F}(x,x,y)\nn\\&=&\frac{a(a+2)}{2(a+1)^2}T_{A\cup B}^{(1,1)}.\label{taba}
\eea 
The last step has been checked numerically for a general set of positive $a$ and $\eta$. Then
\be
T_{A\cup B}^{(m,1)}=[(\frac{\partial}{\partial a})^m(\frac{a(a+2)}{2(a+1)^2})]|_{a=0}\times T_1^{(1,1)}=\frac{(-1)^{m-1}(m+1)!}{2}T_1^{(1,1)}.
\ee 
We will discuss this point later.  

\section{Two dimensional conformal field theory}
We find the exact  correlator $T_{A\cup B}(a,b)$ in two dimensional massless free scalar theory by quantizing it in finite region. In this section, we will study the same correlator for more general two dimensional conformal field theories. Any two dimensional conformal field theory has a general sector which is realized by operator product expansion (OPE) of stress tensor 
\be
T(y)T(y')\sim \frac{c/2}{(y-y')^4}+\frac{2T(y')}{(y-y')^2}+\frac{\partial_{y'}T(y')}{y-y'},
\ee 
where the central charge is $c$. For massless free scalar, $c=1$. Two and three point correlation functions of stress tensors are determined by conformal symmetry up to central charge 
\bea
\langle T(y_1)T(y_2)\rangle&=&\frac{c/2}{y_{12}^4},\\
\langle T(y_1)T(y_2)T(y_3)\rangle&=&\frac{c}{y_{12}^2y_{23}^2y_{13}^2}.
\eea 
We defined $y_{ij}=y_i-y_j$ which is the distance between points $y_i$ and $y_j$. Any higher point functions of stress tensor could be fixed by Ward identity \cite{Belavin:1984vu}
\be
\langle \prod_{i=1}^nT(y_i)\rangle=\langle\sum_{i=2}^n T(y_2)\cdots T(y_{i-1})(\frac{c/2}{(y_1-y_i)^4}+\frac{2T(y_i)}{(y_1-y_i)^2}+\frac{\partial_{y_i}T(y_i)}{y_1-y_i})T(y_{i+1})\cdots T(y_n)\rangle.\label{cf}
\ee 
For example, four point function of stress tensor is 
\bea
\langle\prod_{i=1}^4T(y_i)\rangle&=&c^2(\frac{1}{4y_{12}^4y_{34}^4}+\frac{1}{4y_{13}^4y_{24}^4}+\frac{1}{4y_{14}^4y_{23}^4})\nn\\&&+2c(\frac{1}{y_{12}^2y_{23}^2y_{34}^2y_{24}^2}+\frac{1}{y_{13}^2y_{23}^2y_{34}^2y_{24}^2}+\frac{1}{y_{14}^2y_{23}^2y_{34}^2y_{24}^2})\nn\\&&-2c(\frac{1}{y_{12}y_{23}^3y_{24}^2y_{34}^2}+\frac{1}{y_{12}y_{23}^2y_{24}^3y_{34}^2}-\frac{1}{y_{13}y_{23}^3y_{24}^2y_{34}^2})\nn\\&&-2c(\frac{1}{y_{13}y_{23}^2y_{24}^2y_{34}^3}-\frac{1}{y_{14}y_{23}^2y_{24}^3y_{34}^2}-\frac{1}{y_{14}y_{23}^2y_{24}^2y_{34}^3}).
\eea 
Therefore $T_{A\cup B}(a,b)$ should be completely fixed by conformal symmetry. The modular Hamiltonian in region $A$ is 
\bea
\hat{H}_A&=&2\pi \int_{z_A-R_A}^{z_A+R_A} dz \frac{R_A^2-(z-z_A)^2}{2R_A}T_{tt}(z)\nn\\&=&-\frac{1}{2}\int_{-1}^1 dy (1-y^2)T(y+z_A).
\eea
At the second step, we have used the convention\cite{cft} $T_{tt}=-2\pi T$ and changed variable $y$ to $z$ by
\be
z=R_A(y+z_A).
\ee 
To simplify computation, we already set $R_A=1$.  We can also set $R_B=1$ and $z_B=0$, therefore the modular hamiltonian in region $B$ is 
\be
\hat{H}_B=-\frac{1}{2}\int_{-1}^1 dy (1-y^2)T(y).
\ee 
It is enough to choose the branch $z_A>2$, therefore 
\be
z_A=2\sqrt{1+\frac{1}{\eta}}.
\ee 
The leading term is 
\bea
T_{A\cup B}^{(1,1)}&=&\langle \hat{H}_A\hat{H}_B\rangle_c\nn\\
&=&\frac{c}{8}\int_{-1}^1dy \int_{-1}^1 dy'\frac{ (1-y^2)(1-y'^2)}{(y-y'+z_A)^4}.
\eea
The integral is 
\be
T_{A\cup B}^{(1,1)}=c [-\frac{1}{6}+\frac{(2+\eta)\log(1+\eta)}{12\eta}].
\ee 
For $c=1$, it matches with (\ref{tab11}) which is computed in a rather different way.  Now we compute $T_{A\cup B}^{(2,1)}$, 
\bea
T_{A\cup B}^{(2,1)}&=&\langle \hat{H}_A^2\hat{H}_B\rangle_c\nn\\
&=&:\frac{c}{8}\int_{-1}^1 dy_1\int_{-1}^1dy_2\int_{-1}^1 dy_3\frac{(1-y_1^2)(1-y_2^2)(1-y_3^2)}{(y_1-y_2)^2(y_2-y_3-z_A)^2(y_1-y_3-z_A)^2} :
\eea
$``:\cdots:''$ is understood to remove divergent terms.  There is a pole near $y_1=y_2$ inside the integral, we regularize the integral as if there is no pole\footnote{The regularization is a bit ad hoc at this moment. However, as we will see later, it always recover correct result.}. In practice, one can first do indefinite integral and then taking the limit to the integral bound. If one regularize the integral like this, the integral becomes finite 
\be
T_{A\cup B}^{(2,1)}=c[\frac{1}{2}-\frac{(2+\eta)\log(1+\eta)}{4\eta}].
\ee
It matches with (\ref{tab21}) for $c=1$. This is also a consistency check for the way to regularize the integral. To convince ourselves further, we compute 
\bea
T_{A\cup B}^{(3,1)}&=&\langle \hat{H}_A^3\hat{H}_B\rangle_c\nn\\&=&:\frac{1}{16}\prod_{i=1}^4 \int_{-1}^1 dy_i (1-y_i^2)\times \langle T(y_1+z_A)T(y_2+z_A)T(y_3+z_A)T(y_4)\rangle|_{\mathcal{O}(c)}:.
\eea
The $\mathcal{O}(c^2)$ term has been canceled between the terms $\langle \hat{H}_A^3\hat{H}_B\rangle$ and $-3\langle \hat{H}_A^2\rangle\langle \hat{H}_A\hat{H}_B\rangle$. This can be checked by using the Ward identity for $T(y_4)$.  Using the same method to regularize the integral, we find
\be
T_{A\cup B}^{(3,1)}=c[-2+\frac{(2+\eta)\log(1+\eta)}{\eta}].
\ee
Again, it matches with corresponding scalar result. Finally, 
\bea
T_{A\cup B}^{(2,2)}&=&\langle \hat{H}_A^2\hat{H}_B^2\rangle_c\nn\\
&=&:\frac{1}{16}\prod_{i=1}^4\int_{-1}^1 dy_i (1-y_i^2)\times T(y_1+z_A)T(y_2+z_A)T(y_3)T(y_4)\rangle|_{\mathcal{O}(c)}:\nn\\
&=&c\{\frac{1+\eta}{\eta^2}[4\text{Li}_3(1+\eta)-2\log(1+\eta)\text{Li}_2(1+\eta)+\frac{2\log(1+\eta)}{3}\text{Li}_2(-\eta)+\frac{1+\eta}{3}\log^2(1+\eta)\nn\\&&-\frac{\pi^2}{3}\log(1+\eta)-4\zeta(3)]+\frac{2+\eta}{3\eta}[2\text{Li}_2(-\eta)+3\log(1+\eta)]-\frac{4}{3}\},\label{tab22exa}
\eea  
where the polylogrithm $\text{Li}_n(z)$ is 
\be
\text{Li}_n(z)=\sum_{k=1}^\infty\frac{z^k}{k^n}.
\ee 
In the large distance limit, we find 
\bea
T_{A\cup B}^{(2,2)}&=&c[\frac{\eta^2}{8} - \frac{\eta^3}{8} + \frac{37 \eta^4}{324} - \frac{67 \eta^5}{648} + \frac{
4061 \eta^6}{43200}+\mathcal{O}(\eta^7)]
\eea 
which is exactly (\ref{larget22}) for $c=1$. One can check (\ref{tab22exa}) is indeed (\ref{tab22}) numerically. This result is also mathematically nontrivial since it is a multiple integral of product of hypergeometric functions. It is quite interesting to obtain even higher point correlators of modular Hamiltonian from position space, the result can be expressed as $\text{Li}_n$ functions similar to $T_{A\cup B}^{(2,2)}$. However, the computation becomes cumbersome quickly. 

We will comment on the general structure of $T_{A\cup B}^{(m,n)}$. As we show explicitly, all $\mathcal{O}(c^2)$ terms are canceled  in the correlator of $\langle H_A^mH_B^n\rangle_c$ for $m+n=4,m\ge1,n\ge1$. The property may still be true  for any $m\ge1,n\ge1$ ,
\be
T_{A\cup B}^{(m,n)}=c \ T(m,n;\eta),\label{TABmngen}
\ee
where $T(m,n;\eta)$ is independent of central charge. Technically any $\mathcal{O}(c^k),k\ge 2$ terms are from the most singular term of Ward identity, but these terms are canceled in connected correlation functions
\be
\langle \prod_{i=1}^n T(y_i)\rangle_c=\sum_{i=2}^n\langle  T(y_2)\cdots T(y_{i-1})(\frac{2 T(y_i)}{(y_1-y_i)^2}+\frac{\partial_{y_i}T(y_i)}{y_1-y_i})T(y_{i+1})\cdots T(y_n)\rangle_c.
\ee
Therefore, only  terms proportional to $c$ are left which will lead to (\ref{TABmngen}). One can also understand the property in another way. The modular Hamiltonian is $\mathcal{O}(c)$ for conformal field theory, therefore in the large $c$ limit, 
\be
\langle e^{-a \hat{H}_A-b \hat{H}_B}\rangle\sim e^{-c f}
\ee 
where $f$ is an unkown function which is $\mathcal{O}(c^0)$ at this moment. Taking the logrithmic, one should have 
\be
T_{A\cup B}(a,b)\sim \mathcal{O}(c).
\ee 
Any higher order terms $\mathcal{O}(c^k),\ k\ge2$ must be absent. From Ward identity, the correlators of stress tensor is at least $\mathcal{O}(c)$, therefore 
\be
T_{A\cup B}(a,b)\propto c.
\ee  
For $c=1$, the result should match with the result of two dimensional massless free scalar. Then $T_{A\cup B}(a,b)$ should be 
\be
T_{A\cup B}(a,b)=-\frac{c}{2}\log\det[\bm{1}-\left(\begin{array}{cc}\mathcal{A}&\mathcal{C}\\\mathcal{D}&\mathcal{B}\end{array}\right)],\label{generalTAB}
\ee 
where the matrices $\mathcal{A},\mathcal{B},\mathcal{C}$ and $\mathcal{D}$ are exactly those in massless free scalar theory. 
\section{Operator product expansion of reduced density matrix}
In this section, we will study the relation between operator product expansion of $e^{-a \hat{H}_A}\equiv\rho_A^a$ and the generator $T_{A\cup B}(a,b)$ defined in this paper. We will focus on two dimensional conformal field theory. By definition,
\be
T_{A\cup B}(a,b)=\log\frac{\langle \rho_A^a\rho_B^b\rangle}{\langle\rho_A^a\rangle\langle\rho_B^b\rangle}.
\ee 
We find the correlator in previous sections. However, one can also use operator product expansion to evaluate the same quantity. Notice that $\rho_A^a$ is a nonlocal operator in region $A$, it should be decomposed as a summation of complete orthogonal operators in region $A$ with proper coefficients. Schematically, it is 
\be
\frac{\rho_A^a}{\langle \rho_A^a\rangle}=1+\sum_{\mathcal{O}}c_{\mathcal{O}}(a)(\mathcal{O}+\text{decendants})
\ee 
The terms in $``\text{decendants}"$ should be fixed by conformal symmetry for each primary operators $\mathcal{O}$. Formally expanding $\rho_A^a$ with the powers of modular Hamiltonian, operator product expansion of stress tensor tells us that $\mathcal{O}$ can either be stress tensor or (quasi-)primary operators constructed from multiple stress tensors. The operator with lowest conformal weight is just the stress tensor, therefore 
\be
\frac{\rho_A^a}{\langle \rho_A^a\rangle}=1+c_T(a)\hat{H}_A+\sum_{\mathcal{O}'}c_{\mathcal{O}'}(a)(\mathcal{O}'+\text{decendants}),\label{opeA}
\ee  
where $\mathcal{O}'$ means primary operators whose conformal dimension is at least 4. We have seperated  a term which is related to stress tensor. We expect that the most natural way to organize stress tensor and its decendants in region $A$ is the modular Hamiltonian. To fix the coefficient $c_T(a)$, we compute correlator $T_{A\cup B}(a,b)$ in the large distance limit, 
\bea
T_{A\cup B}(a,b)&=&\log \langle \frac{\rho_A^a}{\langle\rho_A^a\rangle}\frac{\rho_B^b}{\langle\rho_B^b\rangle}\rangle
\nn\\&=&\log\langle (1+c_T(a)\hat{H}_A+\cdots)(1+c_T(b)\hat{H}_B+\cdots)\nn\\&=&c_T(a)c_T(b)\langle \hat{H}_A\hat{H}_B\rangle+\cdots
\eea 
At the second line, we used operator product expansion (\ref{opeA}) for region $A$ and $B$. The $``\cdots"$ terms are subleading terms in the large distance limit since they are contributed by operators with higher conformal weight. The quantity $\langle \hat{H}_A\hat{H}_B\rangle$ has been discussed in previous section, we just borrow the result in large distance limit,
\be
T_{A\cup B}(a,b)=\frac{c}{72}c_T(a)c_T(b)\eta^2+\mathcal{O}(\eta^3)
\ee 
This should match with (\ref{T1ablarge}) since it is the leading term in $T_{A\cup B}(a,b)$, therefore 
\be
c_T(a)=-\frac{a(a+2)}{2(1+a)^2}.\label{cTa}
\ee 
There is no phase factor since $c_T(a)$ is real. Note it seems that 
\be
c_T(a)=\frac{a(a+2)}{2(1+a)^2}
\ee 
also satisfies all the conditions. However, one can expand $\rho_A^a$ for small a, therefore 
\be
\frac{\rho_A^a}{\langle\rho_A^a\rangle}=1-a \hat{H}_A+\mathcal{O}(a^2).
\ee 
This fixes $c_T(a)$ to be (\ref{cTa}) completely. Then
\bea
T_{A\cup B}(a)&=&\frac{\partial}{\partial b}T_{A\cup B}(a,b)|_{b=0}=-\frac{\langle \hat{H}_B \rho_A^a\rangle}{\langle \rho_A^a\rangle}\nn\\&=&-\langle \hat{H}_B(1+c_T(a)\hat{H}_A+\cdots)\rangle\nn\\
&=&\frac{a(a+2)}{2(a+1)^2}T_{A\cup B}^{(1,1)}.
\eea 
This interprets the novel integral property (\ref{taba}). At the second line, we used operator product expansion of $\rho_A^a$. At the last step, we used the property that the vacuum correlation function of any two primary operators with different conformal weight is zero.  To fix the coefficients before primary operator with higher conformal weight in (\ref{opeA}), we should expand the exact result of $T_{A\cup B}(a,b)$ order by order. We leave it for future work.
\section{Comments on holographic dual}
In the context of $AdS_3/CFT_2$, Newton constant of $AdS_3$ gravity is mapped to the central charge of a two dimensional conformal field theory living in the boundary \cite{Brown:1986nw}, 
\be
c=\frac{3\ell}{2G_N},
\ee 
where $\ell$ is $AdS$ radius and $G_N$ is Newton constant. A quantity which is proportional to central charge in conformal field theory will be said to be classical from the gravity side. The correlator $T_{A\cup B}(a,b)$ defined in this work is proportional to central charge for general conformal field theory.This indicates the gravity dual should be a quantity evaluates on a classical configuration.  The construction of the explicit quantity is beyond the scope of this paper.  However, from discussion in the $CFT$ side, there is no $\mathcal{O}(1)$ correction for the correlator, so the quantity will be free from any quantum corrections. 
This will constrain the correlation functions of operators in the bulk through (\ref{holoH}). A similar bulk quantity $T_{A\cup B}(a,b)$ can have a $1/G_N\propto c$ expansion, however, only terms proportional to $c$ will be left and any other correlation functions should be canceled exactly. In higher dimensions, there will be no similar universal result of $T_{A\cup B}(a,b)$ for general conformal field theory, however, we still expect it will constrain bulk correlation functions for any consistent quantum gravity.
\section{Conclusion and discussion}
We evaluate the exact generator $T_{A\cup B}(a,b)$ for two dimensional massless free scalar theory. The result only depends on a set of Bogoliubov matrices. 
We also obtain an exact generator in two dimensional conformal field theories by noticing $T_{A\cup B}(a,b)$ is proportional to the central charge $c$ and matching it with massless free scalar theory.  We could check the result up to $m+n=4$ for general two dimensional conformal field theories.   Higher point correlation functions could be found in principle though the computation will quickly  become messy.  As a by product, we find several exact definite integrals of multiple specific hypergeometric functions. We could check these integrals numerically but a rigorous proof is still lacking.

Our work shows that one can extract finite result from correlation functions $\langle \hat{H}_A^m\hat{H}_B^n\rangle$, though the correlators themselves are divergent in general. The finite part, $\langle\hat{H}_A^m\hat{H}^n_B\rangle_c$ are functions of cross ratio for conformal field theories. The generator $T_{A\cup B}(a,b)$ may be meaningful by itself since its form is formally similar to a correlator of Wilson loops \cite{Maldacena:1998im, Rey:1998ik}.  Actually, given any non-local quantities $e^{Q_A}$ and $e^{Q'_B}$ , one can always define similar correlators. $T_{A\cup B}$ is special as modular Hamiltonian should exist for almost any subregion QFTs.  On the other hand, Wilson loops (or other non-local operators) can only be defined in limited QFTs (e.g. gauge theories).   $T_{A\cup B}$ could be extended to curved spacetimes and may  characterize non-local properties of spacetime. One simple example is shock wave spacetime \cite{tHooft:1987vrq, Aichelburg:1970dh}. The quantization of a massless scalar field in supertranslated shock wave spacetime is studied by \cite{Compere:2019rof}. Assuming $A$ is in the left region of shock wave spacetime while $B$ is in the right region, $A$($B$) is a flat subregion so modular Hamiltonian $\hat{H}_A$($\hat{H}_B$) can be defined as in Minkowski spacetime. However, the correlator $T_{A\cup B}$ is affected by shock wave between $A$ and $B$. In this simple example, subregion observer in $A$($B$) will detect the same phenomenon as in Minkowski spacetime while non-trivial information is stored in  correlators between $A$ and $B$. 

Motivated by the similarity to \cite{Berenstein:1998ij},  We briefly discussed the operator product expansion of $e^{-a \hat{H}_A}$ using the knowledge of $T_{A\cup B}(a,b)$. A complete discussion of operator product expansion may be  possible. We expect to return to this project in the near future. 

In higher dimensions, $T_{A\cup B}(a,b)$ depends on theory and shape of subregions. However, it is still possible to extract finite result in some simple examples. The technics of quantizing field theory in a subregion of Minkowski spacetime developed in this work could be extended to higher dimensions.  

\section*{Acknowledgments} 
The research of J.L. was supported by the
Ministry of Science, ICT $\&$ Future Planning, Gyeongsangbuk-do and Pohang City.
\appendix
\section{Matrix identity}
In this section, we will prove (\ref{ezkh}) in detail. We first note the following identity 
\be
(\bm{K}\bm{H})^n=\bm{K}\bm{S}^\dagger \bm{k}(\bm{k}\bm{\Lambda})^n\bm{S},\quad \forall n\ge1. \label{khn}
\ee
This can be proven by iteration. For $k=1$, (\ref{khn}) is trivially satisfied by using the definition of $\bm{H}$ and $\bm{k}^2=\bm{1}$. Now we assume (\ref{khn}) is valid for any $n_0\ge 1$, then 
\be
(\bm{K}\bm{H})^{n_0+1}=(\bm{K}\bm{H})^{n_0}(\bm{K}\bm{H})=\bm{K}\bm{S}^\dagger \bm{k}(\bm{k}\bm{\Lambda})^{n_0}\bm{S} \bm{K}\bm{S}^\dagger \bm{\Lambda}\bm{S}=\bm{K}\bm{S}^{\dagger}\bm{k}(\bm{k}\bm{\Lambda})^{n_0+1}\bm{S},
\ee 
where at the last step, we used the identity (\ref{sksk}). This proves the identity (\ref{khn}). Now it is easy to find 
\be
e^{-z\bm{K}\bm{H}}=\sum_{n=0}^\infty \frac{(-z)^n}{n!}(\bm{K}\bm{H})^n=1+\sum_{n=1}^\infty\frac{(-z)^n}{n!}\bm{K}\bm{S}^\dagger\bm{k}(\bm{k}\bm{\Lambda})^n\bm{S}=1+\bm{K}\bm{S}^\dagger\bm{k}(e^{-z\bm{k}\bm{\Lambda}}-1)\bm{S}.
\ee 
\section{Normal ordering and parameter differentiation method}
The normal ordering of an operator which is constructed from $a_i,a_i^\dagger$ is to remove all creation operators to the left hand side of annihilation  operators, without considering commutation relations (\ref{annia}). We will use the symbol $\mathcal{N}$ to denote normal ordering. For example 
\be
\mathcal{N}(a_ia_j^\dagger)=a_j^\dagger a_i.
\ee 
We will assume \cite{Wilcox:1967zz}
\be
e^{z\hat{H}}=\mathcal{N}(e^{\sum_{ij}(F_{ij}(z)a_i^\dagger a_j^\dagger+G_{ij}(z)a_ia_j+H_{ij}(z)a_i^\dagger a_j)+K(z)}),\label{ezH}
\ee 
where $F_{ij},G_{ij},H_{ij}$ and $K$ are functions of $z$. It is easy to convince oneself that 
\be
\langle0_M|e^{z\hat{H}}|0_M\rangle=e^{K(z)}
\ee 
since normal ordering will remove any annihilation operator to the right hand side and annihilation operator will annihilate vacuum $|0_M\rangle$. Only $e^K$ is left since it is a number. The differential method is to taking the derivative of $z$ of both sides of (\ref{ezH}), then using the definition of normal ordering to find 
\be
\hat{H}=\sum_{ij}(F_{ij}'a_i^\dagger a_j^\dagger+H_{ij}'a_i^\dagger S(a_j)+G_{ij}'S(a_ia_j))+K',\label{diff}
\ee
where $'$ means the derivative of $z$, $S(a_i)$ is similarity transformation of $a_i$ 
\be
S(a_i)=e^{z\hat{H}}a_i e^{-z \hat{H}}=(e^{-z\bm{K}\bm{H}}\vec{A})_i=\bm{\Theta}_{ij}a_j+\bm{\Phi}_{ij}a_j^\dagger.
\ee 
Similarly, the similarity transformation of $a_ia_j$ is 
\be
S(a_ia_j)=S(a_i)S(a_j)
\ee  
which is a quadratic polynomial of $a_i$ and $a_i^\dagger$. By matching the coefficients before $a_i^\dagger a_j^\dagger, a_i^\dagger a_j,a_i a_j$ and identity of (\ref{diff}), we find an equation set
\bea
\frac{1}{2}\bm{g}^*&=&\bm{\Theta}^T\bm{G}'\bm{\Theta},\label{diffG}\\
-\frac{1}{2}\text{tr}\bm{X}&=&K'+\text{tr}\bm{\Theta}^T\bm{G}'\bm{\Phi}-\frac{1}{2}\text{tr}\bm{h}^*, \label{diffK}
\eea
where we have omitted the differential equations for $F_{ij}$ and $H_{ij}$ since they are not relevant in this work. 
$\bm{G}$ are matrix whose elements are 
\be
\bm{G}_{ij}=G_{ij}.
\ee 
The matrices $\bm{h}$ and $\bm{g}$ are defined as 
\be
\bm{H}\equiv\left(\begin{array}{cc}\bm{h}&\bm{g}\\\bm{g}^*&\bm{h}^*\end{array}\right)
\ee 
Combining with (\ref{bmH}), we have 
\be
\bm{h}=\bm{\alpha}^T\bm{X}\bm{\alpha}^*+\bm{\beta}^\dagger\bm{X}\bm{\beta},\quad \bm{g}=-\bm{\alpha}^T\bm{X}\bm{\beta}^*-\bm{\beta}^\dagger\bm{X}\bm{\alpha}.
\ee 
So $\bm{h}$ is Hermitian and $\bm{g}$ is symmetric,
\be
\bm{h}=\bm{h}^\dagger,\quad \bm{g}=\bm{g}^T.
\ee 
Now we can take z derivative of the matrix $e^{-z\bm{K}\bm{H}}$ and using the definition of $\bm{\Theta}, \bm{\Phi},\bm{h}$ and $\bm{g}$, 
we find 
\be
\left(\begin{array}{cc}\bm{\Theta}'&\bm{\Phi}'\\\bm{\tilde{\Phi}}'&\bm{\tilde{\Theta}}'\end{array}\right)=\left(\begin{array}{cc}\bm{\Theta}&\bm{\Phi}\\\bm{\tilde{\Phi}}&\bm{\tilde{\Theta}}\end{array}\right)\left(\begin{array}{cc}-\bm{1}&0\\0&\bm{1}\end{array}\right)\left(\begin{array}{cc}\bm{h}&\bm{g}\\\bm{g}^*&\bm{h}^*\end{array}\right).
\ee
The differential equation of $\bm{\Theta}$ is 
\be
\bm{\Theta}'=-\bm{\Theta}\bm{h}+\bm{\Phi}\bm{g}^*.\label{difftheta}
\ee 
The differential equation of $K$ is then 
\be
K'(z)=-\frac{1}{2}\text{tr}\bm{X}+\frac{1}{2}\text{tr}\bm{h}^*-\frac{1}{2}\text{tr}\bm{\Theta}^{-1}\bm{\Phi}\bm{g}^*=-\frac{1}{2}\text{tr}\bm{X}-\frac{1}{2}\text{tr}\bm{\Theta}^{-1}\bm{\Theta}'.\label{diffK2}
\ee 
At the  last step, we used the equation (\ref{difftheta}) and 
\be
\text{tr}\bm{h}=\text{tr}\bm{h}^*
\ee 
since $\bm{h}$ is Hermitian. The equation (\ref{diffK2}) can be integrated out explicitly with the initial condition $K(0)=0$, 
\be
K(z)=-\frac{1}{2}z\ \text{tr}\bm{X}-\frac{1}{2}\log\det\bm{\Theta}.
\ee 
This proves the identity (\ref{exp}). 
\section{Bogoliubov Matrices}
The Bogoliubov matrices used in this work are only quadratic in terms of $\bm{\alpha},\bm{\beta}$. 
Depending on the region, Bogoliubov matrices are classified into two classes. The first class is 
\be
\bm{\beta}_A^*\bm{\beta}_A^T,\quad \bm{\beta}_A\bm{\beta}_A^\dagger,\quad \bm{\alpha}_A^* \bm{\beta}_A^\dagger,\quad \bm{\beta}_A\bm{\alpha}_A^T
\ee  
or 
\be
\bm{\beta}_B^*\bm{\beta}_B^T,\quad \bm{\beta}_B\bm{\beta}_B^\dagger,\quad \bm{\alpha}_B^* \bm{\beta}_B^\dagger,\quad \bm{\beta}_B\bm{\alpha}_B^T.
\ee 
Any matrix belongs to first class is constructed from only one region ($A$ or $B$). The second class is 
\bea
&&\bm{\beta}_A^*\bm{\beta}_B^T,\quad\bm{\beta}_A\bm{\beta}_B^\dagger,\quad
\bm{\beta}_A\bm{\alpha}^T_B,\quad\bm{\alpha}_A^*\bm{\beta}_B^\dagger,\\
&&\bm{\beta}_B^*\bm{\beta}_A^T,\quad\bm{\beta}_B\bm{\beta}^\dagger_A,\quad\bm{\beta}_B\bm{\alpha}^T_A,\quad\bm{\alpha}_B^*\bm{\beta}^\dagger_A.
\eea 
Any matrix belongs to second class is constructed from two regions ($A$ and $B$).
We first study matrices in first class. Since region $B$ is similar to region $A$, we can focus on region $A$.  Notice that
\be
\bm{\beta}_A\bm{\beta}_A^\dagger=(\bm{\beta}_A^*\bm{\beta}_A^T)^*,\quad \bm{\beta}_A\bm{\alpha}_A^T=(\bm{\alpha}_A^*\bm{\beta}_A^\dagger)^\dagger,
\ee 
it is enough to consider $ \bm{\beta}_A^*\bm{\beta}_A^T$ and $\bm{\alpha}_A^*\bm{\beta}_A^\dagger$. 
\bea
(\bm{\beta}_A^*\bm{\beta}_A^T)_{vv'}&=&\frac{1}{4\pi^2\sqrt{vv'}}R_A^2\int_0^\infty d\omega \int_{-1}^1 ds \int_{-1}^1 ds' \omega e^{i\omega R_A(s-s')}(\frac{1+s}{1-s})^{ivR_A}(\frac{1+s'}{1-s'})^{-iv'R_A}\nn\\
&=&-\frac{1}{4\pi^2\sqrt{vv'}}\int_{-1}^1 ds\int_{-1}^1 ds' (\frac{1+s}{1-s})^{ivR_A}(\frac{1+s'}{1-s'})^{-iv'R_A}\frac{1}{(s-s'+i\epsilon)^2}\nn\\
&=&-\frac{1}{4\pi^2\sqrt{vv'}}\int_{-\infty}^\infty dt \int_{-\infty}^\infty dt'\frac{e^{2i(tv-t'v')R_A}}{\sinh^2(t-t'+i\epsilon)}\nn\\
&=&-\frac{1}{4\pi^2\sqrt{vv'}}(-1)\frac{4\pi v R_A}{e^{2\pi R_A v}-1}\int_{-\infty}^\infty dt' e^{2iR_A(v-v')t'}\nn\\
&=&\frac{\delta(v-v')}{e^{2\pi R_A v}-1}.
\eea 
At the second step, we inserted a positive imaginary part in the exponential to make the integral finite. At the third step, we changed the varaible $s,s'$ to $t,t'$ by 
\be
s=\tanh t,\quad s'=\tanh t'.
\ee 
Then  residue theorem has been used for the integral of t. The proof of (\ref{asbs}) is similar, we will find a term which is proportional to $\delta(v+v')$. However, since $v$ and $v'$ are assumed to be positive, $\delta(v+v')$ is always zero,
\be
(\bm{\alpha}_A^*\bm{\beta}_A^\dagger)_{vv'}=0.
\ee 
The computation of matrices in second class is quite similar, we will just show the details for $\bm{\beta}_A^*\bm{\beta}_B^T$.
\bea
(\bm{\beta}_A^*\bm{\beta}_B^T)_{v\tilde{v}}&=&\frac{R_AR_B}{4\pi^2\sqrt{v\tilde{v}}}\int_0^\infty d\omega \int_{-1}^1 ds \int_{-1}^1 d\tilde{s}(\frac{1+s}{1-s})^{ivR_A}(\frac{1+\tilde{s}}{1-\tilde{s}})^{-i\tilde{v}R_B} \omega e^{i\omega(z_B-z_A+R_A s-R_B \tilde{s})}\nn\\&=&-\frac{R_AR_B}{4\pi^2\sqrt{v\tilde{v}}}\int_{-1}^1 ds \int_{-1}^1 d\tilde{s}(\frac{1+s}{1-s})^{ivR_A}(\frac{1+\tilde{s}}{1-\tilde{s}})^{-i\tilde{v}R_B}\frac{1}{(z_B-z_A+R_As-R_B\tilde{s})^2}\nn\\&=&-\frac{R_AR_B}{\pi^2\sqrt{v\tilde{v}}}\int_0^\infty dt\int_0^\infty d\tilde{t} \frac{t^{ivR_A}\tilde{t}^{-i\tilde{v}R_B}}{(x_{24}t\tilde{t}+x_{23}t+x_{14}\tilde{t}+x_{13})^2}\nn\\&=&-\frac{R_AR_B}{\pi^2 \sqrt{v\tilde{v}}}B(1+ivR_A,1-ivR_A)B(1+i\tilde{v}R_B,1-i\tilde{v}R_B)x_{23}^{-(1+ivR_A)}x_{14}^{-(1-i\tilde{v}R_B)}\nn\\&&\times x_{13}^{ivR_A-i\tilde{v}R_B}{}_2F_1(1+ivR_A,1-i\tilde{v}R_B,2,1-\frac{x_{24}x_{13}}{x_{14}x_{23}}).
\eea
At the third step, we changed variables $s,\tilde{s}$ to $t,\tilde{t}$ by 
\be
t=\frac{1+s}{1-s},\quad \tilde{t}=\frac{1+\tilde{s}}{1-\tilde{s}}.
\ee 
We also used the definition of $x_{ij}$ given in (\ref{xijdef}).At the last step, we used the integral  
\bea
\int_0^\infty dx\int_0^\infty dy \frac{x^{\mu-1}y^{\nu-1}}{(a x y+b x+c y+d)^\rho}=B(\mu,\rho-\mu)B(\nu,\rho-\nu)b^{-\mu}c^{-\nu}d^{\mu+\nu-\rho}{}_2F_1(\mu,\nu,\rho,1-\frac{ad}{bc})\nn\\\label{int}\eea 
for any $\text{Re}(\rho)>\text{Re}(\mu)>0, \text{Re}(\rho)>\text{Re}(\nu)>0,a>0,b>0,c>0,d>0$. This integral can be proved by \cite{tables}
\bea
&&\int_0^\infty dx \frac{x^{\mu-1}}{(1+\beta x)^\rho}=\beta^{-\mu}B(\mu,\rho-\mu),\quad |\text{Arg}(\beta)|<\pi,\text{Re}(\rho)>\text{Re}(\mu)>0.
\eea and 
\bea
&&\int_0^\infty dx x^{\nu-1}(x+\beta)^{-\mu}(x+\gamma)^{-\rho}=\beta^{-\mu}\gamma^{\nu-\rho}B(\nu,\mu-\nu+\rho){}_2F_1(\mu,\nu,\mu+\rho,1-\frac{\gamma}{\beta}),\nn\\
&&|\text{Arg}(\beta)|<\pi,|\text{Arg}(\gamma)|<\pi,\text{Re}(\nu)>0,\text{Re}(\mu)>\text{Re}(\nu-\rho).
\eea 
Beta function is symmetric 
\be
B(x,y)=B(y,x),
\ee 
we also notice the identity 
\be
1-\frac{x_{24}x_{13}}{x_{14}x_{23}}=-\eta,\quad x_{12}=2R_A,\quad x_{34}=2R_B.
\ee 
Therefore we have 
\be
(\bm{\beta}_A^*\bm{\beta}_B^T)_{v\tilde{v}}=G(ivR_A,-i\tilde{v}R_B).
\ee 
All other matrices in second class can be computed in a similar way.

\end{document}